\documentclass[twocolumn]{aastex62}

\received{04-May-2020}
\revised{15-Jun-2020}
\accepted{15-Jun-2020}

\submitjournal{ApJ}

\shortauthors{D\'ek\'any \& Grebel}

\begin{document}

\title{Near-Infrared Search for Fundamental-mode RR~Lyrae Stars Toward the Inner Bulge by Deep Learning}

\correspondingauthor{Istv\'an D\'ek\'any}
\email{dekany@uni-heidelberg.de}

\author[0000-0002-0786-7307]{Istv\'an D\'ek\'any}
\affil{Astronomisches Rechen-Institut, Zentrum f\"ur Astronomie der Universit\"at Heidelberg,
M\"onchhofstr. 12-14, 69120 Heidelberg, Germany}

\author[0000-0002-1891-3794]{Eva K. Grebel}
\affiliation{Astronomisches Rechen-Institut, Zentrum f\"ur Astronomie der Universit\"at Heidelberg,
M\"onchhofstr. 12-14, 69120 Heidelberg, Germany}

\begin{abstract}

Aiming to extend the census of RR~Lyrae stars to highly reddened low-latitude regions of the central Milky Way, we performed a deep near-IR variability search using data from the VISTA Variables in the V\'ia L\'actea (VVV) survey of the bulge, analyzing the photometric time series of over a hundred million point sources. In order to separate fundamental-mode RR~Lyrae (RRab) stars from other periodically variable sources, we trained a deep bidirectional long short-term memory recurrent neural network (RNN) classifier using VVV survey data and catalogs of RRab stars discovered and classified by optical surveys. Our classifier attained a $\sim99\%$ precision and recall for light curves with signal-to-noise ratio above 60, and is comparable to the best-performing classifiers trained on accurate optical data. Using our RNN classifier, we identified over 4300 hitherto unknown bona fide RRab stars toward the inner bulge. We provide their photometric catalog and VVV $J,H,K_s$ photometric time-series.

\end{abstract}


\section{Introduction} \label{sec:introduction}

RR~Lyrae stars play an instrumental role in our understanding of the structure and formation history of our home Galaxy. These pulsating horizontal-branch stars serve as the ``Swiss Army Knives of Astronomy'' \citep{2011rrls.conf..181S}; they are accurate standard candles \citep[e.g.,][]{2018ApJ...869...82R}, which also makes them valuable reddening estimators; they are good photometric tracers of metallicity \citep{1996A&A...312..111J,2018ApJ...857...55H}; and, due to their highly constrained ages and high number density, they act as proxies of the oldest stellar populations \citep[see, e.g.,][and references therein]{2018ApJ...857...54D}. 

The number of known RR~Lyrae stars in the Galactic halo, bulge, and disk increased swiftly in recent years due to the voluminous catalogs yielded by large photometric surveys such as the Catalina Sky Survey \citep{2014ApJS..213....9D}, All-Sky Automated Survey for Supernovae \citep[ASAS-SN,][]{2019MNRAS.486.1907J,2020MNRAS.491...13J}, Panoramic Survey Telescope and Rapid Response System \citep[Pan-STARRS,][]{2017AJ....153..204S}, Gaia \citep{2019A&A...622A..60C}, the Optical Gravitational Lensing Experiment \citep[OGLE,][]{2019AcA....69..321S} and the VISTA Variables in the V\'ia L\'actea survey \citep[VVV,][]{2018ApJ...857...54D}. 

The volume density of RR~Lyrae stars steeply increases toward the Galactic Center, where they reach a projected number density of at least $10^3$ stars per square degrees, and their spherically symmetric distribution in the halo transitions into an oblate spheroid within the inner $\sim$3~kpc, according to data from the OGLE-IV survey \citep{2015ApJ...811..113P}. The orbits of the vast majority of RR~Lyrae stars in the inner Milky Way seem to be confined to the bulge \citep{2019MNRAS.487.3270P,2020arXiv200411382K}, where they have witnessed the earliest formation history of our Galaxy.

RR~Lyrae stars already provided key insight into the properties of the oldest stellar populations of the bulge. Their spatial and kinematical distributions are significantly different from those of the metal-rich intermediate-age stars traced by red clump (RC) giants. While the boxy/peanut distribution \citep{2013MNRAS.435.1874W} and cold kinematics \citep[e.g.,][]{2019MNRAS.489.3519C} of bulge RC stars are consistent with a pseudo-bulge formed and evolved through disk instabilities \citep[e.g.,][]{2012ApJ...756...22N}, RR~Lyrae stars paint a different picture about the old, metal-poor component of the bulge. Their spheroidal, barely elongated distribution \citep[e.g.,][]{2013ApJ...776L..19D,2015ApJ...811..113P} and hot kinematics \citep{2016ApJ...821L..25K,2020arXiv200411382K,2019MNRAS.487.3270P} point toward the possible existence of a classical component in a composite Galactic bulge \citep{2013ApJ...763...26O}. 

In order to gain a deeper understanding of the physical properties of the old and metal-poor bulge component, a full census of RR~Lyrae stars in the inner bulge is required. Former studies of the RR~Lyrae stars' spatial distribution \citep[e.g.,][]{2013ApJ...776L..19D,2015ApJ...811..113P,2019MNRAS.484.4833P} were constrained by the census of these objects being largely limited to the southern half of the bulge where a sufficiently large sample at $b < -2^\circ$ with contiguous celestial coverage was provided by OGLE. Although the new data release of the OGLE-IV survey \citep{2019AcA....69..321S} and the additional objects identified in the Gaia DR2 \citep{2019A&A...622A..60C} significantly improved their census toward the bulge, the completeness of both surveys ends abruptly at low Galactic latitudes. At $\vert b \vert \lesssim 2^\circ$, extinction by interstellar dust concentrated along the Galactic plane pushes the apparent magnitudes of RR~Lyrae stars beyond the faint detection limit of both surveys.

This limitation of optical surveys can be overcome by near-infrared (near-IR) photometry, where the effect of interstellar extinction is greatly diminished. The VVV survey of the Galactic bulge \citep{2010NewA...15..433M}, carried out with the VIRCAM \citep{2006SPIE.6269E..0XD} near-IR imager of the VISTA telescope at the European Southern Observatory on Cerro Paranal, is particularly suitable for extending the census of RR~Lyrae stars toward the inner bulge and also to fill the discontinuities in the OGLE survey's celestial coverage in the northern bulge \citep[see][]{2019AcA....69..321S}. In an earlier study, we employed VVV data to identify RR~Lyrae stars along the southern mid-plane \citep{2018ApJ...857...54D} using machine learning techniques, but the survey's bulge area ($\vert l \vert \lesssim 10^\circ$) was excluded from our study. \citet{2018ApJ...863...79C} published a catalog of 488 new RR~Lyrae candidates in the inner bulge based on VVV data, but their analysis was constrained to within a $100'$ radius around the Galactic Center.

In this paper, we leverage VVV photometry to perform a deep near-IR search for fundamental-mode RR~Lyrae (RRab) stars toward the bulge, with the aim of complementing their earlier census, in particular, extending it to $\vert b \vert \lesssim 2^\circ$ where optical surveys are hindered by extreme reddening. In order to provide accurate light-curve classifications, we employ deep learning, using a recurrent neural network. This paper is structured as follows. In Sect.~\ref{sec:observations}, we give an overview of the VVV photometry, its calibration, and a general variability search. In Sect.~\ref{sec:classification}, we discuss the problem of near-IR light-curve classification, briefly review recurrent neural network architectures, and describe our data representation, as well as the procedure of model selection, training, validation, and the estimation of the classifier's performance. The deployment of the classifier and the resulting catalog of new RR~Lyrae stars is described in Sect.~\ref{sec:catalog}. We summarize our results in Sect.~\ref{sec:summary}.

\section{Observations, variability analysis}\label{sec:observations}

We analyzed near-IR photometric time-series of the VVV survey \citep{2010NewA...15..433M}, acquired between 2010 and 2015. Our target area encompasses the inner bulge region at low Galactic latitudes consisting of the VVV fields b299--b304 and b308--b382 \citep[see][for field definitions]{2010NewA...15..433M}, spanning across $-10^\circ \lesssim l \lesssim +10.5^\circ$ and $-2.5^\circ \lesssim b \lesssim +3.8^\circ$.
Every field was observed at 100 epochs in the $K_s$ band and at two epochs in the $J$ and $H$ bands of the VISTA photometric system with a non-uniform cadence, except for the fields b308--b310, which have several hundred epochs in $K_s$. The limiting apparent magnitudes are highly space-varying depending on source density, ranging up to $\sim16.5$\,mag in the $K_s$, and $\sim 20$\,mag in the $J$ band \citep[see also][]{2012A&A...537A.107S}.

This study is based on the standard photometric catalogs created by the VISTA Data Flow System \citep[VDFS,][]{2004SPIE.5493..401E} provided by the Cambridge Astronomy Survey Unit (CASU). The pipeline for image processing and aperture photometry is described by \citet{2004SPIE.5493..411I}. We relied on data products derived from single detector frame stacks, a.k.a. {\em pawprints}. Depending on an object's position on the VIRCAM array, each photometric epoch consists of 1--6 individual {\em pawprint} measurements within a $\sim3$-minute interval. The source tables from these measurements were used for deriving a unified catalog following the positional cross-matching procedure discussed by \citet{2018ApJ...857...54D}. Our target area contains over $10^8$ point sources.

Before further processing, the photometric zero-points (ZPs) of all data were re-calibrated following the procedures of \citet{zpcalib}. This treatment eliminated time- and space-varying ZP biases present in the CASU catalogs, attributed mostly to the blending of photometric calibrator stars in the 2MASS \citep{2006AJ....131.1163S} catalog, which the VVV photometry is tied to. The corrected ZP offsets do not only affect the mean magnitudes of objects, but they can severely distort the light curves of variable stars in highly crowded regions, thus hindering their detection and classification (see \citealt{zpcalib} and \citealt{2019ApJ...883...58D} for further details).

We searched for variable stars in the VVV data following the procedures described by \citet{2018ApJ...857...54D}, here we only give a concise overview. Point sources with ostensible light variation were selected on the basis of variability indices that take advantage of the correlated photometric sampling. Roughly $\sim 10\%$ of all objects were selected in this way and were subjected to a search for periodic signals in the $[0.28,0.98]$~day interval employing the Generalized Lomb-Scargle Periodogram \citep[GLS,][]{2009A&A...496..577Z} method. The above interval encompasses the periods of bulge RRab stars previously discovered by the OGLE-IV survey \citep{2019AcA....69..321S}.  Variable star candidates with at least 30 photometric epochs in the $K_s$-band were propagated to automated light-curve classification by a recurrent neural network, which we present in Sect.~\ref{sec:classification}. We emphasize that detecting new overtone RR~Lyrae (RRc) stars is out of our scope due to their very small amplitudes and featureless light curves in the near-IR. Likewise, the number of measurements and the temporal sampling of the VVV photometric time series do not allow us to search for double-mode RR~Lyrae (RRd) stars.

\section{Light-curve classification}\label{sec:classification}

The sheer amount of data from large time-domain photometric surveys such as VVV render automated light-curve classification a mandatory task. While supervised machine-learning is routinely applied to the optical data of variable stars \citep[e.g,][]{2016A&A...587A..18K}, their classification in the near-IR has been more challenging due to the inherently subtler features in their light curves, and the relative scarcity of high-quality training data \citep[see, e.g.,][]{2014A&A...567A.100A}. 

Time-series classification problems in astronomy have been traditionally framed in a `feature-based' approach, whereby a number of {\em features} (e.g., descriptive statistics, parameters of regression models, etc.) are derived from the photometric data, and these are used as descriptive variables at the input of a classification model designed for structured data, such as a random forest classifier \citep[see, e.g.,][and references therein]{2007A&A...475.1159D,2012ApJS..203...32R}. We took a similar approach to develop a near-IR classifier for RR~Lyrae stars \citep{2016A&A...595A..82E} and deployed it on the Galactic disk section of the VVV survey \citep{2018ApJ...857...54D}. 

However, as we discussed in \citet{2019ApJ...883...58D}, such feature-based representation of the data can be sub-optimal if strongly correlated features occupy complicated manifolds in the parameter space. In addition, erratic data distribution can yield biased features, further limiting the classification performance. For example, features derived from a model representation of a periodic light curve (e.g., from the regression of a Fourier series or a Gaussian process) can easily become biased if the phase distribution of the measurements is irregular, e.g., contains a large gap, causing the fit to diverge from the optimal solution. Input features susceptible to bias can also enhance the possible problem of data mismatch, whereby the distributions of the training/validation and target data sets differ, causing biased performance estimates.

In this study, we take on a different approach of light-curve classification by employing a recurrent neural network (RNN), which directly takes the photometric measurements as its input, thus avoiding the aforementioned issues. Sequence models based on RNNs have proven to be an extremely versatile means of accurately solving various problems ranging from time-series classification and forecasting to natural language processing and neural translation \citep[see, e.g., ][and references therein]{2015arXiv150600019L}. Recently, the application of RNNs in astronomy has also started to unfold, e.g., for the classification of supernova light curves \citep{2017ApJ...837L..28C}. \citet{2018NatAs...2..151N} have clearly demonstrated the superiority of RNN-autoencoders over traditional feature-based classification of unevenly sampled variable star light curves in multi-class problems, in terms of performance and scalability.

In the following, we provide a short summary of RNNs and their advantages over classical, fully connected neural networks; and explain the importance and functionality of long-range connections in advanced RNN architectures.

\subsection{Classical and recurrent neural networks}

A classical, fully connected neural network (a.k.a. multi-layer perceptron, MLP) takes a finite vector $\mathbf{x}$ of descriptive variables (features) as its input and propagates it through a series of $L$ hidden layers, where layer $l$ performs a linear and a consecutive non-linear transformation of the following form on its input:

\begin{equation}
\mathbf{a}^{[l]} = g(\mathbf{W}^{[l]}\mathbf{a}^{[l-1]}+\mathbf{b}^{[l]})~.\label{MLP}
\end{equation}

\noindent Here, $\mathbf{W}^{[l]}$ and $\mathbf{b}^{[l]}$ are the weight matrix and bias vector of the $l$-th layer, respectively, i.e., free parameters of the model, and $g$ is a nonlinear activation function \citep[see, e.g.,][for a review]{2018arXiv181103378N}. The layer's output, the so-called activation vector $\mathbf{a}^{[l]}$, forms the input of layer $l+1$, and $\mathbf{a}^{[0]}=\mathbf{x}$.

The prediction is computed in the output layer following the last hidden layer, and its specific form depends on the type of the classification or regression problem to be solved. In case of binary (i.e., two-class) classification, the output layer takes the following form:

\begin{equation}
\hat y  = \sigma(\mathbf{w}^{\rm [out]}\mathbf{a}^{[L]} + b^{\rm [out]})~,\label{sigmoid}
\end{equation}

\noindent where $\hat y$ is the predicted probability of the given example described by $\mathbf{x}$ being of class $y=1$, the $\mathbf{w}^{\rm [out]}$ vector and $b^{\rm [out]}$ scalar are free parameters of the output layer, and $\sigma(x)$ is the sigmoid function, i.e., $\sigma(x) = (1+e^{-x})^{-1}$.

A sufficiently deep ($L\gg1$) MLP with large $\mathbf{W^{[l]}}$ parameter matrices (i.e., many ``neurons'') is a highly flexible function, capable of modeling complicated, highly non-linear interdependencies between $\mathbf{x}$ and $y$. However, the MLP architecture is best suited for structured data, e.g., an input vector $\mathbf{x}$ comprising a set of statistical features derived from, e.g., a light curve.

In principle, we could also directly pass a time series to an MLP in its input vector $\mathbf{x}$, thus considering each time step as an input feature, but this would have significant disadvantages. Firstly, all time series would have to be adjusted to have the same length. More importantly, the MLP architecture is unable to share similar features between its neurons that are learned across different positions in its input sequence. In other terms, if different parts of a light curve have similar shapes, they will be learned by different neurons, i.e., the corresponding model parameters will not be shared within the model. This key capability, i.e., parameter sharing, is a prominent advantage of convolutional and recurrent neural networks.

In our previous study, we used a convolutional neural network (CNN) for the classification of near-IR Cepheid light curves \citep{2019ApJ...883...58D}. Our choice over an RNN was mainly motivated by the heterogeneity and modest size of the training set available for that particular problem. In order to suppress the disrupting effect of significant photometric noise on the learning process (i.e., to avoid that the model ``learns'' noise features in the small training set along with real light curve features), we used a regressed periodic light curve model evaluated on an equidistant grid of phases as the light curve representation at the CNN's input. Since high-quality training examples of near-IR RR~Lyrae light curves are available in large numbers, we opted to rely on RNNs in the present study, which have the most native architectures for sequence classification problems.

Let $\mathbf{x}^{<t>}$ denote the $t$-th time step ($t\in\{1,T_x\}$) of the input sequence (time series). In the basic RNN architecture, the $\mathbf{a}^{[l]<t>}$ activation vector of the $l$-th ($l\in\{1,L\}$) hidden layer at time step $t$ is computed by the following formula:

\begin{equation}
\mathbf{a}^{[l]<t>} = \tanh(\mathbf{W_{aa}}^{[l]}\mathbf{a}^{[l]<t-1>} + \mathbf{W_{ax}}^{[l]}\mathbf{a}^{[l-1]<t>}+\mathbf{b_a}^{[l]})~,
\end{equation}\label{RNN}

\noindent where $\mathbf{W_{aa}}^{[l]}$ and $\mathbf{W_{ax}}^{[l]}$ are the weight matrices and $\mathbf{b_a}^{[l]}$ is the bias vector of the $l$-th layer. We emphasize that these free parameters of the model's $l$-th layer are shared across all time steps. Similarly to an MLP (Eq.~\ref{MLP}), the sequence of $\mathbf{a}^{[l]<t>}$ activation vectors of each hidden layer serves as the input sequence of the next, and the input of the first hidden layer is the sequence $\mathbf{x}$, i.e., $\mathbf{a}^{[0]<t>}=\mathbf{x}^{<t>}$.  All elements of the $\mathbf{a}^{[l]<0>}$ activation vectors are defined to be 0. We note that $\mathbf{x}^{<t>}$ can be multi-dimensional, e.g., its dimensions can carry magnitudes measured in different filters, in case of multi-band photometry.

The stack of $L$ recurrent layers described above can be considered as an encoder that transforms the input sequence $\mathbf{x}^{<t>}$ into the $\mathbf{a}^{[L]<t>}$ vectors of abstract features. In case of a time-series classification problem, the output $\mathbf{a}^{[L]<T_x>}$ of the last layer and at the last time step can be used as the input of an output layer, which in analogy to Eq.~\ref{sigmoid}, has the form:

\begin{equation}
\hat y  = \sigma(\mathbf{w_{ya}}\mathbf{a}^{[L]<T_x>} + b_y)~.\label{rnn-sigmoid}
\end{equation}

\noindent Alternatively, a number of fully connected layers (Eq.~\ref{MLP}) can be included between the last recurrent layer and the output layer, in order to increase the model's complexity.

In case of a binary classification problem, the optimal model parameters are found by minimizing the following cost function for the training set:

\begin{equation}
J = \sum_{i=1}^N \mathcal{L}_i(\hat y_i, y_i) = \sum_{i=1}^N -y_i\log{\hat y_i} - (1-y_i)\log(1-\hat y_i)~,
\end{equation}\label{loss}

\noindent where $\mathcal{L}_i$ is the binary cross-entropy loss, $y_i \in \{0,1\}$ is the true class, and $\hat y_i \in (0,1)$ is the predicted class of the $i$-th training example. Since the partial derivatives of $J$ with respect to the model parameters can be computed explicitly, the model's optimal parameters can be found by a gradient-based minimization algorithm.

A significant limitation of basic RNNs is that they are not efficient at capturing long-range interdependencies between distant elements in the input sequence, i.e., the values of $\mathbf{a}^{<t>}$ are mainly influenced by values of $\mathbf{x^{<t'>}}$, where $t-t'$ is small. The manifestation of this in the network's optimization is that the gradients of $J$ with respect to the activations will exponentially decrease toward the beginning of the sequence. In the literature, this effect is commonly referred to as the vanishing gradient problem \citep[e.g.,][]{1997Neco...9..1735}, and it greatly decreases the effectiveness of basic RNNs on longer sequences. 

To combat vanishing gradients, in their seminal paper \citet{1997Neco...9..1735} proposed a modification of the basic RNN, called the resulting model architecture Long Short-Term Memory (LSTM) network. In order to capture long-range connections in the input sequence, they introduced a memory cell $\mathbf{c^{<t>}}$ in the RNN hidden layer. The first hidden layer performs the following operations on its input vector $\mathbf{x}^{<t>}$ at time step $t$:

\begin{eqnarray}
\mathbf{\widetilde c}^{<t>} &=& \tanh(\mathbf{W_{ca}}\mathbf{a}^{<t-1>} + \mathbf{W_{cx}}\mathbf{x}^{<t>} + \mathbf{b_c}) \label{cell_cand} \\
\mathbf{\Gamma_u}^{<t>} &=& \sigma(\mathbf{W_{ua}}\mathbf{a}^{<t-1>} + \mathbf{W_{ux}}\mathbf{x}^{<t>} + \mathbf{b_u}) \label{update_gate} \\
\mathbf{\Gamma_f}^{<t>} &=& \sigma(\mathbf{W_{fa}}\mathbf{a}^{<t-1>} + \mathbf{W_{fx}}\mathbf{x}^{<t>} + \mathbf{b_f}) \label{forget_gate} \\
\mathbf{c}^{<t>} &=& \mathbf{\Gamma_u} \ast \mathbf{\widetilde c}^{<t>} + \mathbf{\Gamma_f} \ast \mathbf{c}^{<t-1>} \label{cell} \\
\mathbf{\Gamma_o}^{<t>} &=& \sigma(\mathbf{W_{oa}}\mathbf{a}^{<t-1>} + \mathbf{W_{ox}}\mathbf{x}^{<t>} + \mathbf{b_o}) \label{output_gate} \\
\mathbf{a}^{<t>} &=& \mathbf{\Gamma_o} \ast \tanh(\mathbf{c}^{<t>}) \label{activation_lstm}~,
\end{eqnarray}\label{lstm}

\noindent where the `$\ast$' operator denotes element-wise multiplication, and, similarly to the basic RNN, the elements of the various matrices and vectors denoted by $\mathbf{W}$ and $\mathbf{b}$, respectively, are the model's free parameters fitted to the data.

\begin{figure}[!t]
\includegraphics[width=0.48\textwidth]{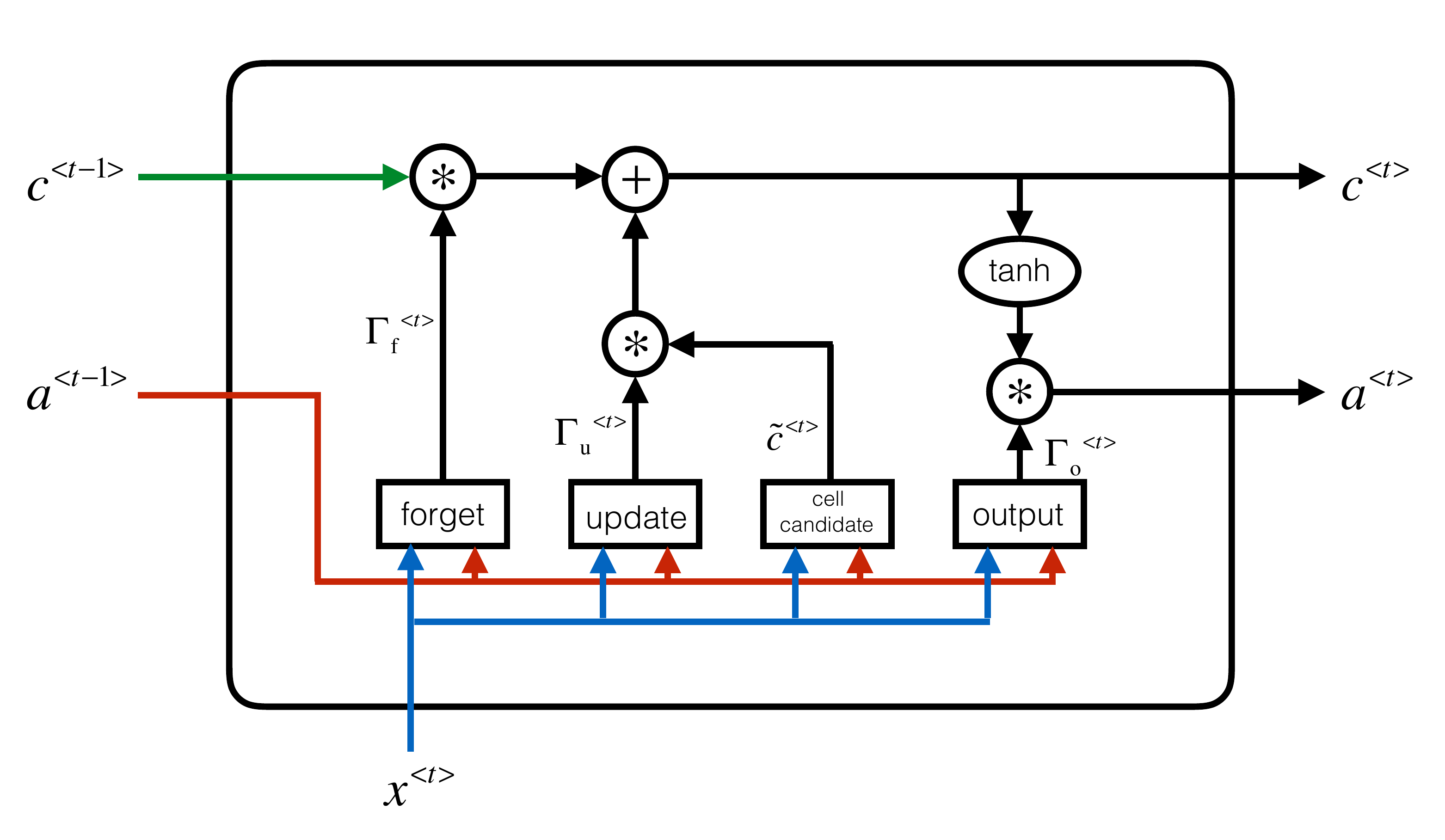}
\caption{Schematic graph showing the functionality of an LSTM unit (see also Eqs.~\ref{cell_cand}--\ref{activation_lstm}).
\label{fig:lstm_cell}}
\end{figure}

At each time step, a memory cell candidate $\mathbf{\widetilde c}^{<t>}$ is computed from the activation vector of the previous time step and the input vector of the current time step (Eq.~\ref{cell_cand}). Moreover, three additional vectors $\mathbf{\Gamma_u}^{<t>}, \mathbf{\Gamma_f}^{<t>}, \mathbf{\Gamma_o}^{<t>}$ are computed, which are called update, forget, and output gates (Eqs.~\ref{update_gate},\ref{forget_gate}, and \ref{output_gate}), respectively. All elements of these can take values between 0 and 1. The update and forget gates determine whether the previous memory cell is replaced by its candidate value (Eq.~\ref{cell}). Finally, the layer's activation is computed from the memory cell and the output gate (Eq.~\ref{activation_lstm}). Figure~\ref{fig:lstm_cell} provides an intuitive graphical summary of the functionality of an LSTM unit.

While Eqs.~\ref{cell_cand}--\ref{activation_lstm} only describe the operations done by the first hidden layer for the sake of a more compact notation, LSTMs can also be stacked, similarly to basic RNNs, i.e., the sequence of activation vectors can be directed to the input of the next hidden layer ({\em cf.} Eq.~\ref{RNN}.)

Since its invention, LSTM networks and its variations \citep[e.g.,][]{2014arXiv1409.1259C,2014arXiv1412.3555C} have proven to be extremely versatile tools for a wide range of machine-learning applications on sequence data \citep[for a recent review, see][]{2019Neco...31..1235}. Recurrent networks, including LSTMs and their variants can have bidirectional architecture in case the entire sequence is available upon input (such as in our case). Each hidden layer of a bidirectional network consists of two effectively separate networks, one processing the input sequence in the original, while the other one does so in the reversed direction, outputting two sequences of activation vectors. These are usually combined or stacked upon output into a single activation sequence, before directing them to  the input of the next hidden layer. Bidirectional networks have more parameters, and provide higher flexibility for the model compared to single-directional networks, and are therefore also easier to overfit.

In addition to the model parameters, neural networks such as LSTMs have several hyper-parameters to be optimized as well. For example, hyper-parameters such as the number of stacked recurrent and (optionally) fully connected layers, and their respective number of ``neurons'' (i.e., the number of rows in the $\mathbf{W}$ parameter matrices) regulate the model's complexity, and optimization algorithms also have their own hyper-parameters. These have fixed values during a model's training, and their optimal values can be searched by  cross-validation (CV). During CV, labeled data are randomly split into training and validation sets one or multiple times. The model is optimized on the training set(s) with its hyper-parameters fixed to some trial values, and then evaluated on the validation set(s) using some type of performance metric. The optimal hyper-parameters are determined by repeating this procedure for various values, and finding those that maximize the performance metric. In case of deep networks, a full grid-search of all hyper-parameters is generally discarded due to the immense computational cost, and instead a limited number of intuition-guided experiments are carried out until a desired performance is achieved on the validation set.

\subsection{Light-curve representation}\label{lc_representation}

All light curves in our training, test, and target sets described in the following Sections were subjected to the iterative regression algorithm described by \citet{2018ApJ...857...54D}. In brief, for each trial aperture we iteratively perform a period search with the GLS method and fit a truncated Fourier sum to the light curve by a robust non-linear regression with outlier rejection, determining the optimal Fourier order by 10-fold cross-validation. The optimal aperture is selected by minimizing the regression loss.

\begin{figure}[!t]
\includegraphics[width=0.48\textwidth]{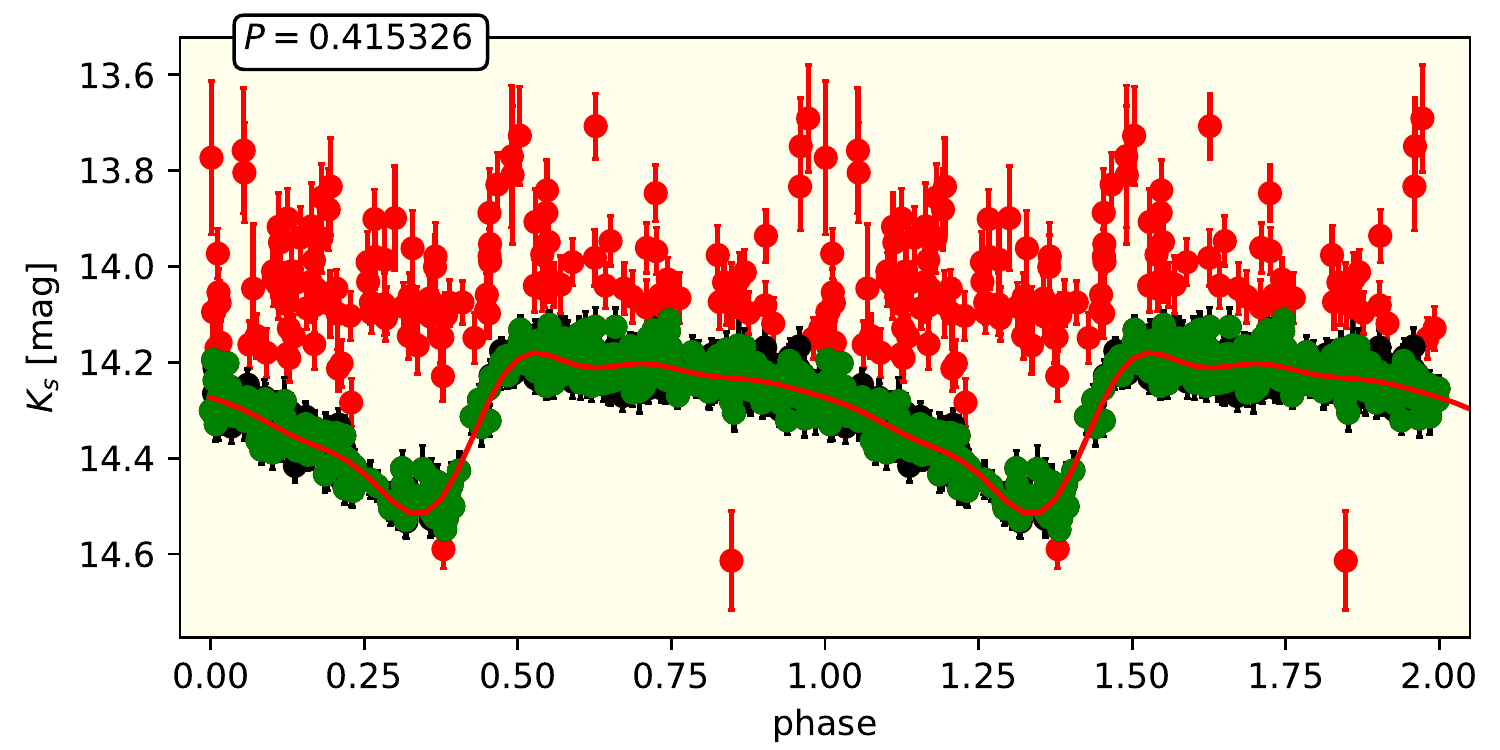}
\caption{The VVV $K_s$-band light curve of the RRab star OGLE-BLG-RRL-12700 phase-folded with the pulsation period. The red points were rejected by our iterative regression algorithm. The black points were kept and binned to one data point per epoch (green symbols). The algorithm found the smallest aperture to be the optimal one. The red curve shows a 5-order truncated Fourier sum fitted to the data.
\label{fig:fit_example}}
\end{figure}

Each light curve is then phase-folded with the optimal period returned by the regression algorithm, as well as phase-aligned by requiring the first Fourier term to be at zero phase. The data are also binned to one point per epoch, i.e., the weighted mean is computed for groups of data points (measured by different chips) that correspond to the same observational epoch (see Sect.~\ref{sec:observations}). Figure~\ref{fig:fit_example} demonstrates the effectiveness of the regression procedure on the light curve of the object OGLE-BLG-RRL-12700, whose VVV photometry is particularly affected by temporal blending with surrounding objects. 

After post-processing, each object could thus be represented by a scalar feature, i.e., the period $P$, acting as a scaling parameter along the temporal dimension; and a two-dimensional sequence:

\begin{eqnarray}
\mathbf{x}^{<t>}&=& \left(\begin{array}{cc}{m^{<t>}} \\ {\phi_P^{<t>}}\end{array} \right),~t=\{1,\dots,N_{\rm ep}\}~,\\
\phi_P(T)&=&{\rm mod}[(T+P\cdot\Phi_1/(2\pi))/P]~,
\end{eqnarray}\label{eq:input_seq}

\noindent where $m^{<t>}$ and $\phi_P^{<t>}$ are the mean-subtracted $K_s$ magnitudes and the corresponding phases of the binned light curve, respectively, $\Phi_1$ is the phase of the first Fourier term, $T$ is the observation time, and $N_{\rm ep}$ is the number of observational epochs. 

\begin{figure*}
\plottwo{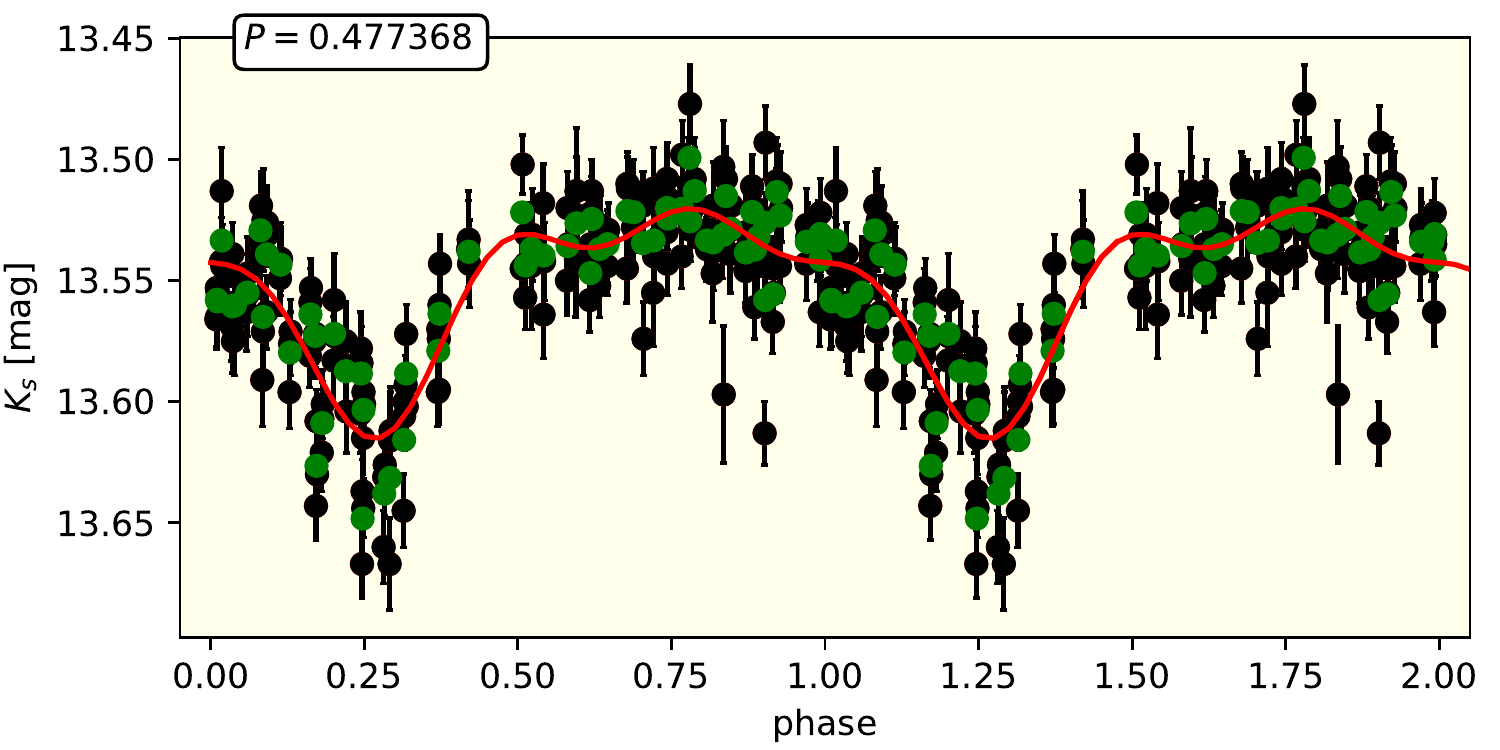}{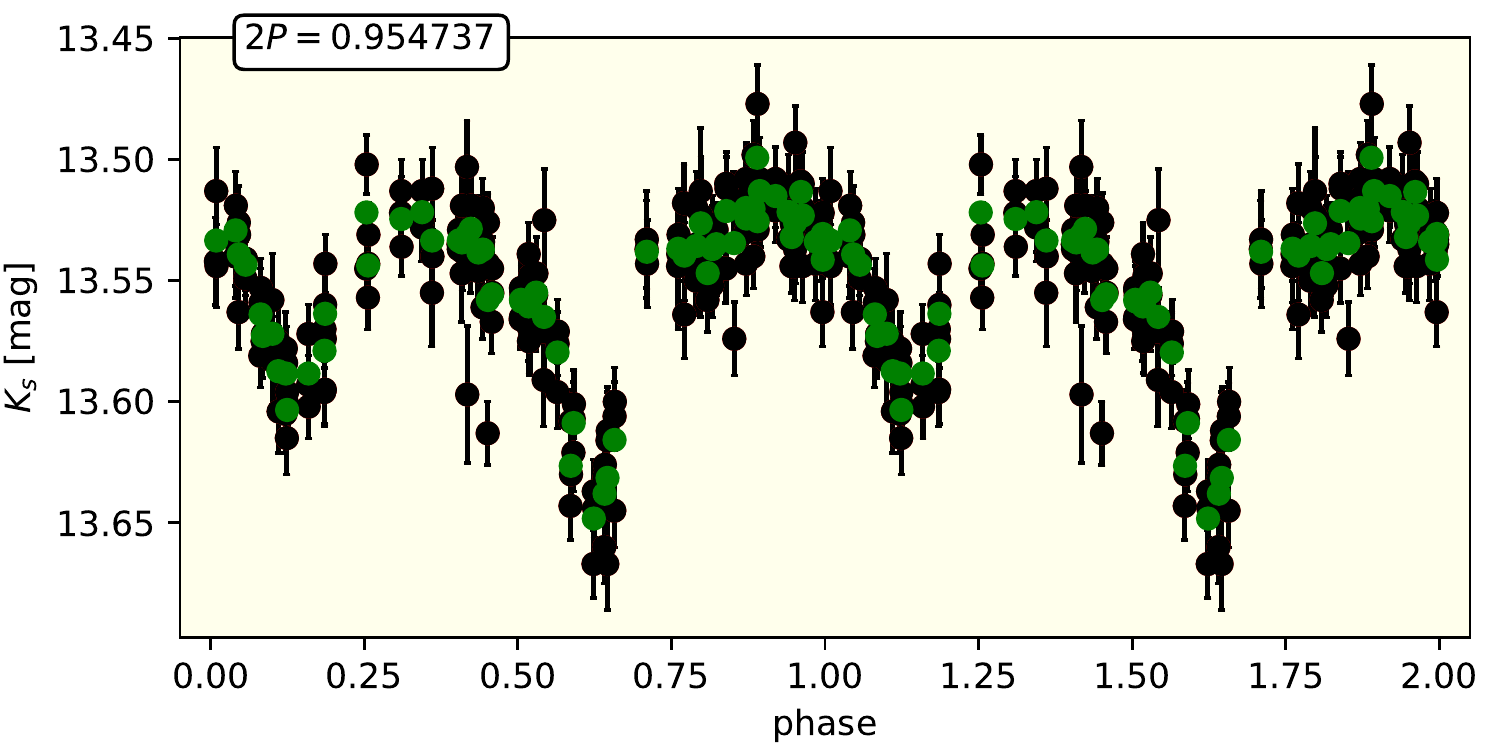}
\caption{
Left: $K_s$-band VVV light curve of an object in our test set, phase-folded with the period returned by our non-linear regression algorithm initialized with the GLS period. Right: the same light curve, phase-folded with twice the period as in the left panel, revealing the alternating minima of the eclipsing binary star.
\label{fig:binary_example}}
\end{figure*}

However, if we use the above data representation as the input of the RNN classifier, we run into the following problem. Our training and test sets (and presumably our target set) contain a large number of contact and semi-detached binary stars, whose light-curves have minima with alternating depth (which often only slightly differ). Period-search algorithms such as the GLS tend to find half of the true period of such objects, thus in the above $\{m^{<t>},\phi_P^{<t>},P\}$ representation, their primary and secondary minima are ``folded'' on top of each other, as illustrated by Fig.~\ref{fig:binary_example}. Since the temporal sampling, and thus the phase relations of the minima differ from object to object, it is extremely difficult for RNNs to learn and distinguish such objects from RRab stars with closely symmetrical light curves, as they regularly fail to distinguish the overlapping minima of binary stars from increased photometric noise. 

We could attack the above problem from a period-search standpoint, by replacing GLS with another method that is more sensitive to light curves with alternating minima, such as the phase dispersion minimization \citep[PDM,][]{1978ApJ...224..953S} algorithm. However, this would only diminish, rather than eliminate the problem, since a non-negligible fraction of eclipsing binaries with undersampled minima and /or minima with sufficiently similar depth would still end up identified with half of their true periods, thus still causing unwanted confusion with certain RRab stars. Instead, we let the RNN itself directly perceive the difference between photometric noise and overlapping light-curve features by adding two additional dimensions to the input sequence, namely the magnitudes $m'$ and phases $\phi_{2P}$ of the (binned) light curve phase-folded with $2P$ (where $P$ is the period found by our algorithm). As we will see in Sect.~\ref{subsec:performance}, this approach effectively eliminates the problem related to the periods of eclipsing binaries, and leads to outstanding classification performance.

Finally, we emphasize that our data representation significantly differs from the one used for the multi-class RNN light-curve classifier by \citet{2018NatAs...2..151N}, and thus corresponds to a rather different framing of a similar problem. In their model, the input sequence of the RNN consists of a vector of (one or more) magnitudes and corresponding observation times, thus framing the classification problem in an end-to-end approach. In other words, periods are not explicitly computed, but the frequency-domain properties of the input sequences are inferred by the RNN itself (but without explicitly returning the periods of variable stars). While this is a remarkable capability of the RNNs, we opted not to follow this path in addressing our binary classification problem for the following reasons. Firstly, GLS periods are inexpensive to compute for VVV data due to the relatively low number of data points per light curve, and the corresponding regression procedure is also necessary to select the optimal aperture for each star, as well as to omit strong outliers (see Fig.~\ref{fig:fit_example}). Secondly, and most importantly, RRab stars are {\em linearly separable} from many other variable star types only based on their periods. Thus, our classification problem is largely simplified by the exclusion of all sources from our training and target sets that have aperiodic light curves or show periodic signals outside the period range of known RRab stars. Finally, we are explicitly interested in the periods of the objects in our target set, thus we would have to compute them anyway.

\subsection{Training set}\label{subsec:training_set}

In order to assemble a labeled data set for training and cross-validation, we relied entirely on objects in the OGLE-IV collection of RR~Lyrae stars in the bulge \citep{2014AcA....64..177S}, also covered by the VVV survey. The two main advantages of this dataset are that it consists of VVV photometry, thus the distribution of the training and target sets will be as similar as possible; and since its classification labels are based on high-quality optical photometry, the misclassification rate of OGLE RRab stars is negligible \citep{2019AcA....69..321S}.

We collected the ZP-corrected VVV $K_s$-band photometry of the objects that had the best positional cross-match within a radius of $1''$ of the $27,480$ RRab stars in the OGLE-IV catalog published by \citet{2014AcA....64..177S}. The light curves were processed according to the procedure described in Sect.~\ref{lc_representation} and by keeping their periods fixed to the OGLE-IV values. In order for a good tradeoff between training data quality and training set size, we included those light curves in our training set that passed the following selection criteria:

\begin{eqnarray}
&&S/N = A_{K_s}\sqrt{N_{\rm ep}} / \sigma \geq 60~;\\
&&N_{\rm ep} \geq 40~;\\
&&C_P \geq 0.8~;\\
&&C_{2P} \geq 0.8~;\\
&&12 \leq \langle K_s \rangle \leq 15.5~;\\
&&0.28\,{\rm d} \leq P \leq 0.98\,{\rm d}~; \label{period_sel}
\end{eqnarray}\label{training_set_criteria}

\noindent where $S/N$ is an estimate of the signal-to-noise ratio\footnote{We will use this definition of $S/N$ throughout the paper.};  $A_{K_s}$ and $\sigma$ are the total (peak-to-valley) amplitude and the residual standard deviation of the Fourier fit, respectively; and $C_P$ is the phase coverage\footnote{$1-$maximum phase lag} of the VVV $K_s$ time series corresponding to the period $P$. This was followed by a further quality check by visual inspection, whereby we rejected a further 144 light curves due to various problems, typically blending, and/or undersampled rising branch.

The application of further selection criteria on the {\em optical} light-curve quality of the training set was deemed unnecessary due to the different sensitivity ranges of OGLE and VVV and the rather different apparent brightnesses of the objects in the two surveys due to interstellar extinction. Moreover, we emphasize that we did not exclude any objects based on their intrinsic properties and/or peculiarity because we wanted to avoid our classifier to favor certain subtypes of RRab stars. Likewise, we were aiming to obtain classification performance estimates for the entire RRab population mixture observed towards the bulge.

A total of $15,964$ RRab (i.e., $y=1$) light curves were thus selected for the training set. We note that during our analysis, the OGLE-IV collection of RR~Lyrae stars was updated \citep{2019AcA....69..321S}, thus complementing the \citet{2014AcA....64..177S} sample by several thousand new objects in the VVV survey's bulge area. We used the VVV light curves of these newly identified OGLE RRab stars to create an explicit {\em test set} for measuring the classifier's performance (see Sect.~\ref{subsec:performance}).

The training set of non-RRab (i.e., $y=0$) light curves was compiled from the catalog of periodic variable star candidates identified in the VVV data according to Sect.~\ref{sec:observations} in the same period range as the RRab stars (Eq.~\ref{period_sel}). 
By this approach, we naturally include already known variable stars of various types, such as eclipsing binaries and anomalous Cepheids from OGLE \citep{2017AcA....67..297S,2016AcA....66..405S}, which fall into our detectability range and selection criteria. In addition, our approach has important advantages over relying solely on the VVV data of non-RRab variable stars that were previously known from overlapping surveys. Firstly, we also include other types of variable objects that have remained undetected or unpublished by other surveys, but are nevertheless present in our datasets. Secondly, by applying the same detection and selection procedure for the training, test, and target sets, we ensure that the data distribution within them is as similar as possible, which is a key to good performance on the target set. For example, if our pipeline tends to yield biased periods for certain types of objects (e.g., eclipsing binaries, as discussed in Sect.~\ref{lc_representation}) in the target set, then such biases will also be represented by the training set in the same way, thus our model can learn them. Lastly, our training set selection enables our model to learn erratic periodic light curves that arise from various instrumental effects endemic to the VVV survey, and which also contaminate our target set.

First, we selected the complement set of our variable star catalog with respect to the OGLE-IV RRab sample. The purity of the resulting non-RRab sample depends on the completeness of the OGLE-IV catalog. On the one hand, it is desirable to collect the non-RRab training examples from low-latitude regions with high point-source density in order for them to have a similar data distribution as the target set in the inner bulge. On the other hand, the completeness of the OGLE-IV RRab sample falls with decreasing latitude, due to the increasingly limiting effect of interstellar extinction on the source detection ability of the OGLE survey. In order to have a good balance between the two effects, we collected non-RRab light curves of sources from an area encompassing approximately $18^\circ\times4^\circ$ of the southern bulge, comprising the VVV fields b245--b295 and b305--b307. These data were processed and selected similarly to the training RRab stars, resulting in a total of $22,288$ non-RRab ($y=0$) light curves for our training set.

\begin{figure}
\includegraphics[width=0.5\textwidth]{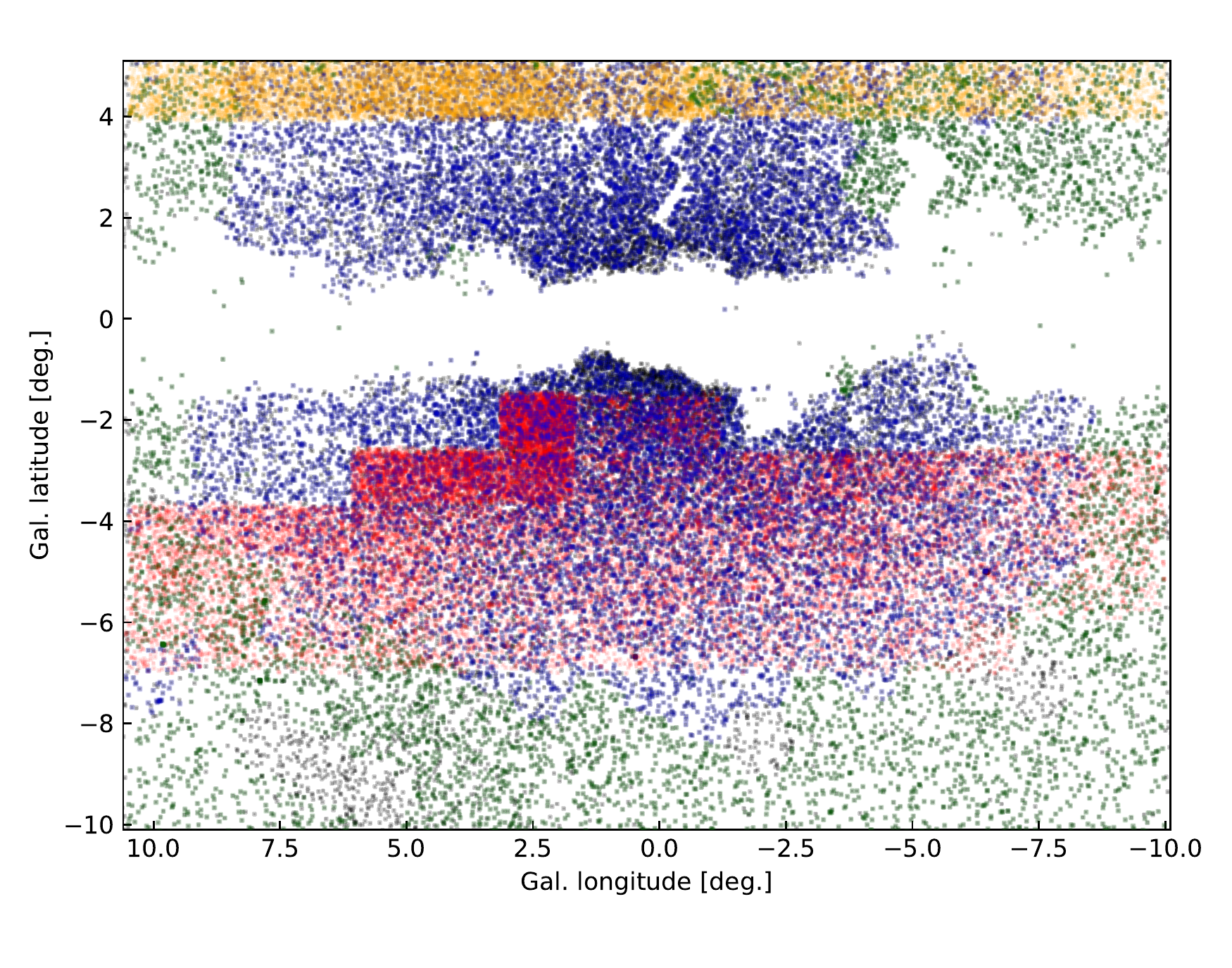}
\caption{Celestial distribution of the objects in our training and test sets in Galactic coordinates. Blue: RRab training set, red: non-RRab training set, green: RRab test set, orange: non-RRab test set. Black points show OGLE-IV RRab stars that were included in neither the training, nor the test set.
\label{fig:lb_map}}
\end{figure}

The celestial distribution of the objects in our training set is shown in Fig.~\ref{fig:lb_map} with red and blue dots. We note that the higher density of non-RRab training objects close to Baade's window ($2^\circ \lesssim l \lesssim 6^\circ$, $-4^\circ \lesssim b \lesssim -2^\circ$) is due to the much denser temporal sampling of the VVV survey in this area (up to several hundred epochs per light curve), which allowed a higher detection rate of variable stars.

\subsection{Model selection and optimization}

\begin{figure}
\includegraphics[width=0.48\textwidth]{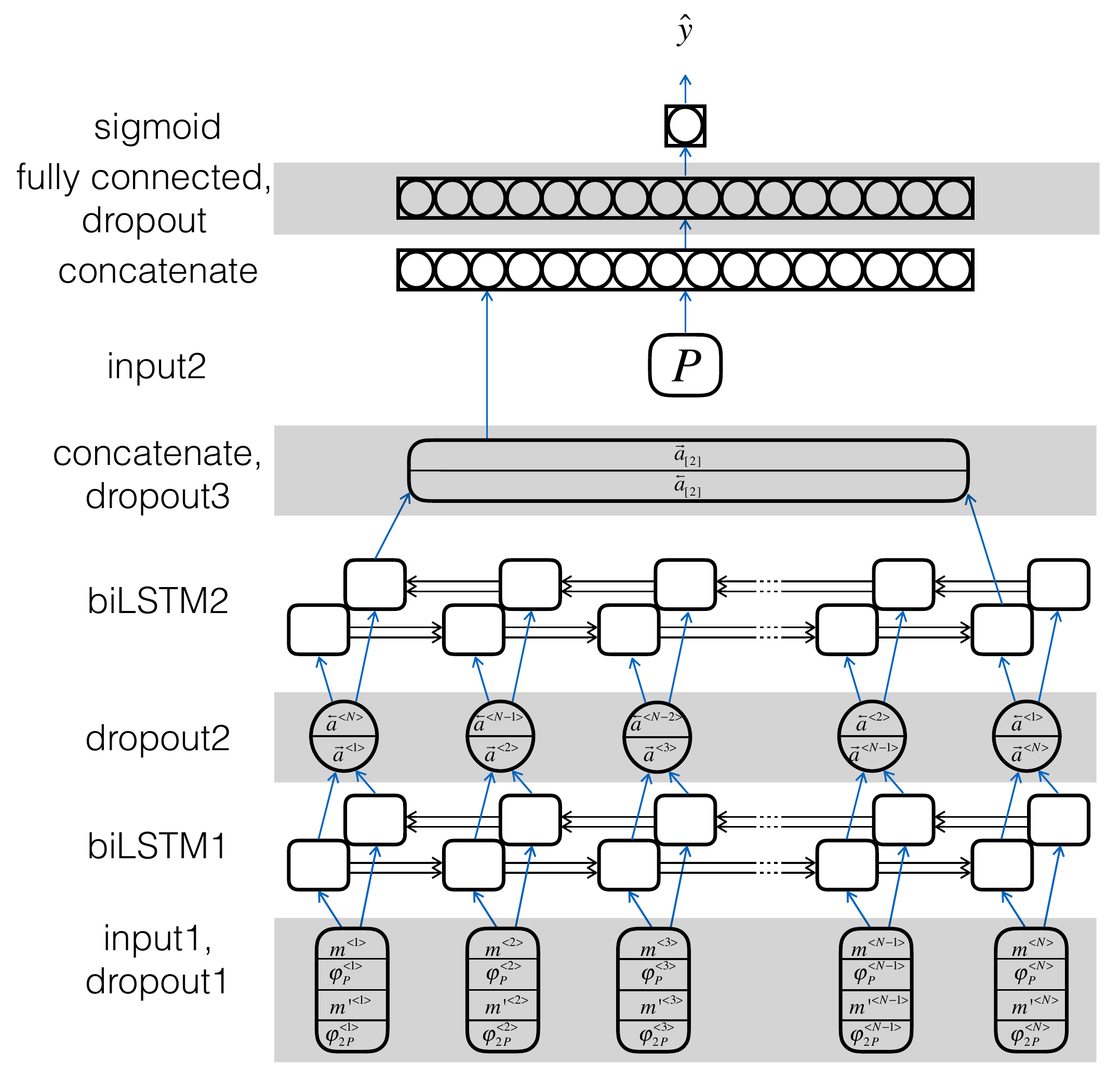}
\caption{Schematic graph of the bidirectional LSTM network used for RRab light-curve classification in this study. Each of the empty boxes in the biLSTM1 and biLSTM2 layers represent an LSTM unit outlined in Fig.~\ref{fig:lstm_cell}.
\label{fig:lstm_network}}
\end{figure}

The architecture of our LSTM-based RNN classifier is summarized by Fig.~\ref{fig:lstm_network}. The network takes the 4-dimensional sequence $\{ m^{<t>},\phi_P^{<t>},m'^{<t>},\phi_{2P}^{<t>} \}$ on its first input layer and feeds it into a bidirectional LSTM (biLSTM) layer. This is optionally followed by a second biLSTM layer. The forward and backward activation vectors on the first recurrent layer's output are concatenated and subsequently dropout \citep{dropoutpaper} is applied on the sequence. Dropout has a regularizing effect by randomly dropping network units (in this case, time steps) at training, hence preventing the RNN from overfitting the data. We can also apply dropout on the first input layer.

The forward and backward activation vectors of the last and first units of the last biLSTM layer are concatenated and flattened into a one-dimensional vector, then dropout is applied once more on the elements of this vector. Afterwards, the activations are concatenated with the second input layer, containing the additional scalar feature, i.e., the period. The resulting vector is fed into a fully connected layer (Eq.~\ref{MLP}), followed by another dropout, and finally the output layer with sigmoid activation function (Eq.~\ref{rnn-sigmoid}).

Our RNN has several hyper-parameters that govern the model's complexity, such as: the number of biLSTM layers, the number of neurons in the recurrent and fully connected layers, the dropout rate at various stages, the activation function of the fully connected layer, etc. We experimented with several variants of this network architecture by tuning these hyper-parameters. The models have been implemented and trained using the TensorFlow \citep{2016arXiv160304467A} and Keras \citep{keras} application programming interfaces.

Each model (corresponding to a fixed combination of hyper-parameters) was trained using the Adam optimization algorithm \citep{2014arXiv1412.6980K} with a mini-batch size of $256$, an initial learning rate of $0.0015$ and a learning rate decay of $8\cdot10^{-5}$. We iterated Adam through 300 training epochs\footnote{At each training epoch, $N_{\rm tr}/N_{\rm mb}$ iterations are performed, where $N_{\rm tr}$ is the number of training examples, and $N_{\rm mb}=256$ is the mini-batch size.}, where a good convergence was reached for every variant of the model. The models were evaluated via 5-fold cross-validation by their standard classification accuracy, which is a good performance metric, since our training set is well-balanced between the two classes. In each fold, we randomly split the data into training and validation sets with a ratio of $1:0.15$. 

\begin{deluxetable}{lll}
\tablecaption{Properties of our RNN classifier\label{tab:bestrnn}}
\tablehead{
\colhead{Layer} & \colhead{hyper-params.} & \colhead{num. of params.}
}
\startdata
dropout1 & rate=0.05 & 0 \\
biLSTM1 & 48 neurons & 20352 \\
dropout2 & rate=0.15 & 0 \\
biLSTM2 & 32 neurons & 33024 \\
dropout3 & rate=0.15 & 0 \\
fully connected & 64 neurons & 4224 \\
dropout3 & rate=0.5 & 0 \\
sigmoid & 1 neuron & 65
\enddata
\end{deluxetable}

\begin{figure*}
\plottwo{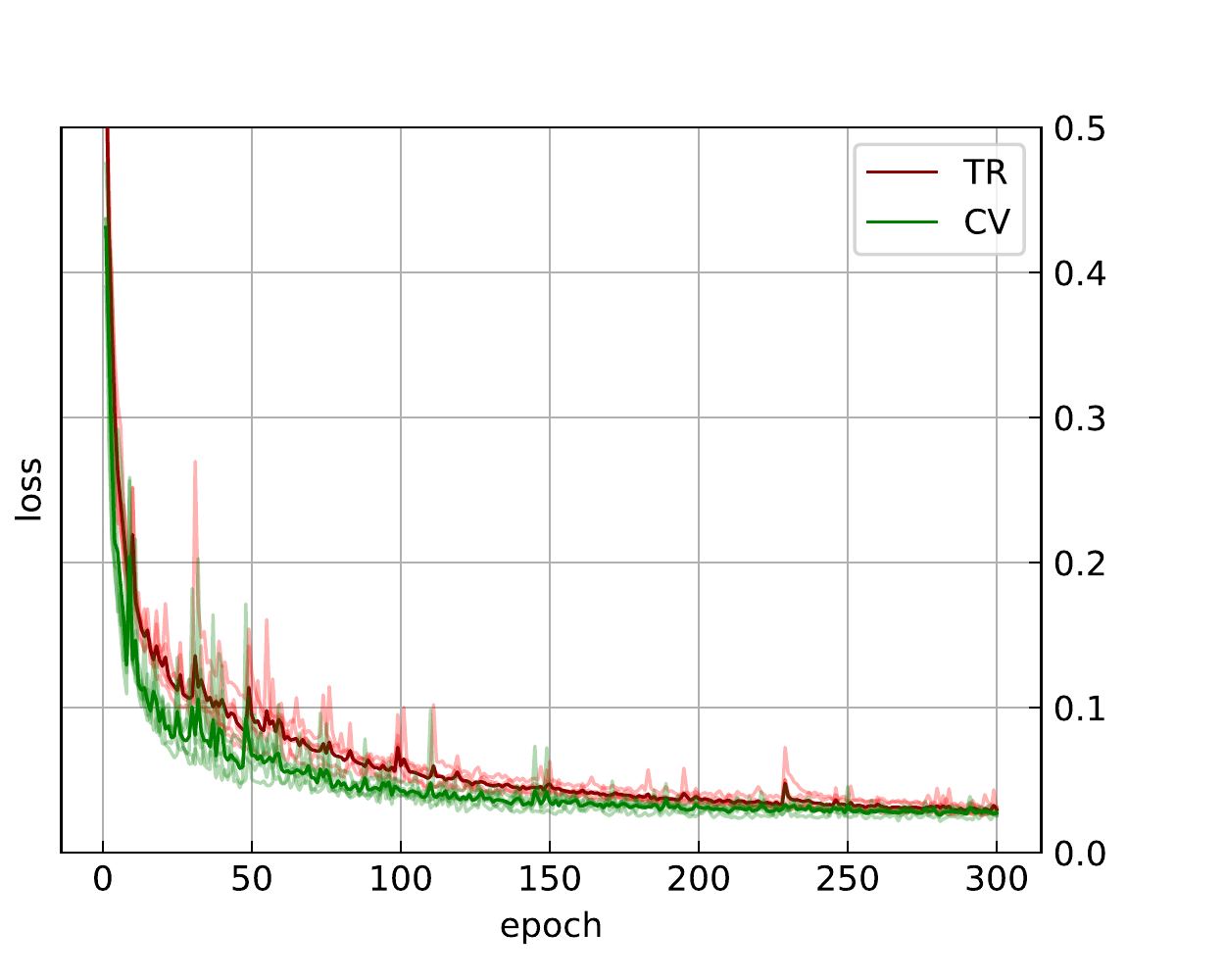}{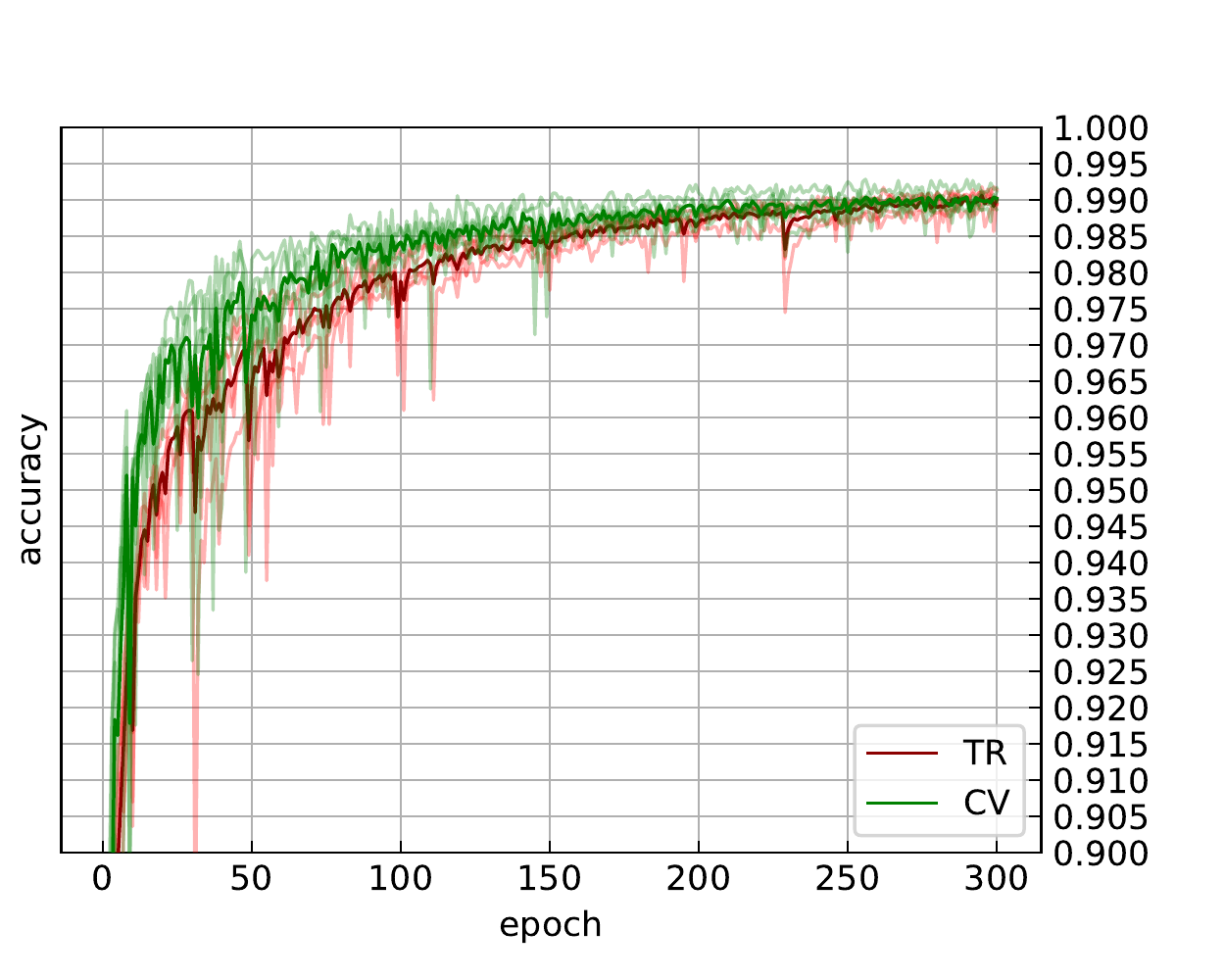}
\caption{
Left: training (TR, red) and mean cross-validation (CV, green) loss as a function of training epochs. Semi-transparent curves show values for each fold separately, while thick lines show their mean. Right: classification accuracy as a function of training epoch, using the same notation.
\label{fig:loss_acc}}
\end{figure*}

The main properties of our best-performing model are summarized by Table~\ref{tab:bestrnn}. We used the ReLu activation function \citep{2018arXiv181103378N} in the fully connected layer, two biLSTM layers, and applied dropout at the first input layer, after each biLSTM layer, and at the fully connected layer. Figure~\ref{fig:loss_acc} shows the binary cross-entropy loss and the classification accuracy of the best model as a function of the training epoch, measured on the training and validation sets. The learning curves converge to virtually the same asymptotes for both the training and validation data, showing an excellent bias-variance tradeoff. We note that the model being more accurate on the validation set at early stages of the training is an effect of using dropout regularization. At training, dropout sets a percentage of various units (sequence steps and neurons) of the RNN to zero, while at testing time (i.e., at the evaluation on the validation set), the full network is used, thus the model is more robust. This effect gradually diminishes as the model becomes more skilled, having passed several training epochs.

After training the model with the best-performing architecture, we performed a visual inspection of the light curves that were misclassified either in the training set or at cross-validation. We detected a relatively small number of objects that could be visually classified as RRab stars with very high certainty, but were labeled as non-RRab stars, i.e., are missing from the OGLE-IV catalog, and are not listed in any other variable star catalog. This can be either because they are too faint in the optical $I$ band, or they lie in narrow discontinuities in the celestial coverage of the OGLE survey \citep[see][]{2019AcA....69..321S}, or they were simply misclassified by the OGLE-IV survey. Given the uncertainty of their visual classification based on near-IR data, we excluded them from the training set, and re-trained our classifier on the remaining data. We repeated this procedure in 3 iterations. A total of 79 objects were omitted from the training set in this way, which are likely newly discovered RRab stars.

Finally, the model was retrained on the entire cleaned training data set, i.e., without cross-validation, and the performance of the resulting final classifier was measured on an explicit test data set.

\subsection{Classification performance}\label{subsec:performance}

The performance of our best model was measured by 5-fold cross-validation using four different standard metrics, namely classification accuracy ($\mathcal{A}$), precision ($\mathcal{P}$), recall ($\mathcal{R}$), and the $F_1$-measure, defined as follows:

\begin{eqnarray}\label{eq:metrics}
\mathcal{A} &=& (N_{tp} + N_{tn}) / (N_{tp} + N_{tn} + N_{fp} + N_{fn}) \\
\mathcal{P} &=& N_{tp}/(N_{tp}+N_{fp}) \\
\mathcal{R} &=& N_{tp}/(N_{tp}+N_{fn}) \\
F_1 &=& 2\mathcal{P}\mathcal{R}/(\mathcal{P}+\mathcal{R}).
\end{eqnarray}

\noindent Here, $N_{tp}$, $N_{tn}$, $N_{fp}$, and $N_{fn}$ denote the number of true positives, true negatives, false positives, and false negatives, respectively, where an object classified as RRab is defined to be a positive example. The resulting values for these metrics are summarized by Table~\ref{tab:perf} (column `CV').

\begin{deluxetable}{lcc}
\tablecaption{Performance metrics for our best model measured by cross-validation (CV) and on an explicit test set.\label{tab:perf}}
\tablehead{
\colhead{Metric} & \multicolumn2c{Value}
}
\startdata
\hline
& CV & test \\
& & ($S/N>60$) \\
\hline
accuracy & 0.990 & 0.983 \\
precision & 0.985 & 0.986 \\
recall & 0.992 & 0.992 \\
$F_1$ & 0.988 & 0.989 \\
\enddata
\end{deluxetable}

In order to obtain an unbiased estimate of the classifier on our target dataset, we evaluated our best model on a labeled explicit test dataset, which was completely unseen by the model during training and cross-validation. The RRab light curves in this test set comprise VVV data of those stars in the latest OGLE-IV catalog by \citet{2019AcA....69..321S} that were not included in previous OGLE data releases (Fig.~\ref{fig:lb_map}). The light curves of periodic non-RRab variable stars for the test set were obtained in a similar way as for the training set: we took the complement set of the variable stars identified in the VVV data with respect to the OGLE-IV catalog in an area where the completeness of the OGLE survey is very high (and is disjunct from the area used for the training set). For this purpose, we used an approximately $1^\circ$-wide stripe between Galactic longitudes of $ -10^\circ \lesssim l \lesssim 10.5^\circ$, namely the VVV fields b383--b396 (see Fig.~\ref{fig:lb_map}, orange dots). The test data were processed according to Sect.~\ref{lc_representation} in the same way as those in the training set, except that we included all light curves with $S/N>30$, $N_{\rm ep.}\geq30$, and $\langle K_s \rangle \geq 12$\,mag, resulting in 4876 RRab and 7853 non-RRab time series. 

A visual inspection of the light curves that were misclassified by our RNN revealed 23 objects that are most probably RRab stars included neither in the OGLE-IV nor in the Gaia DR2 RR~Lyrae catalogs, thus probably had wrong labels in the test set. We did not use these stars in the final performance evaluation of our classifier.

\begin{figure}
\includegraphics[width=0.48\textwidth]{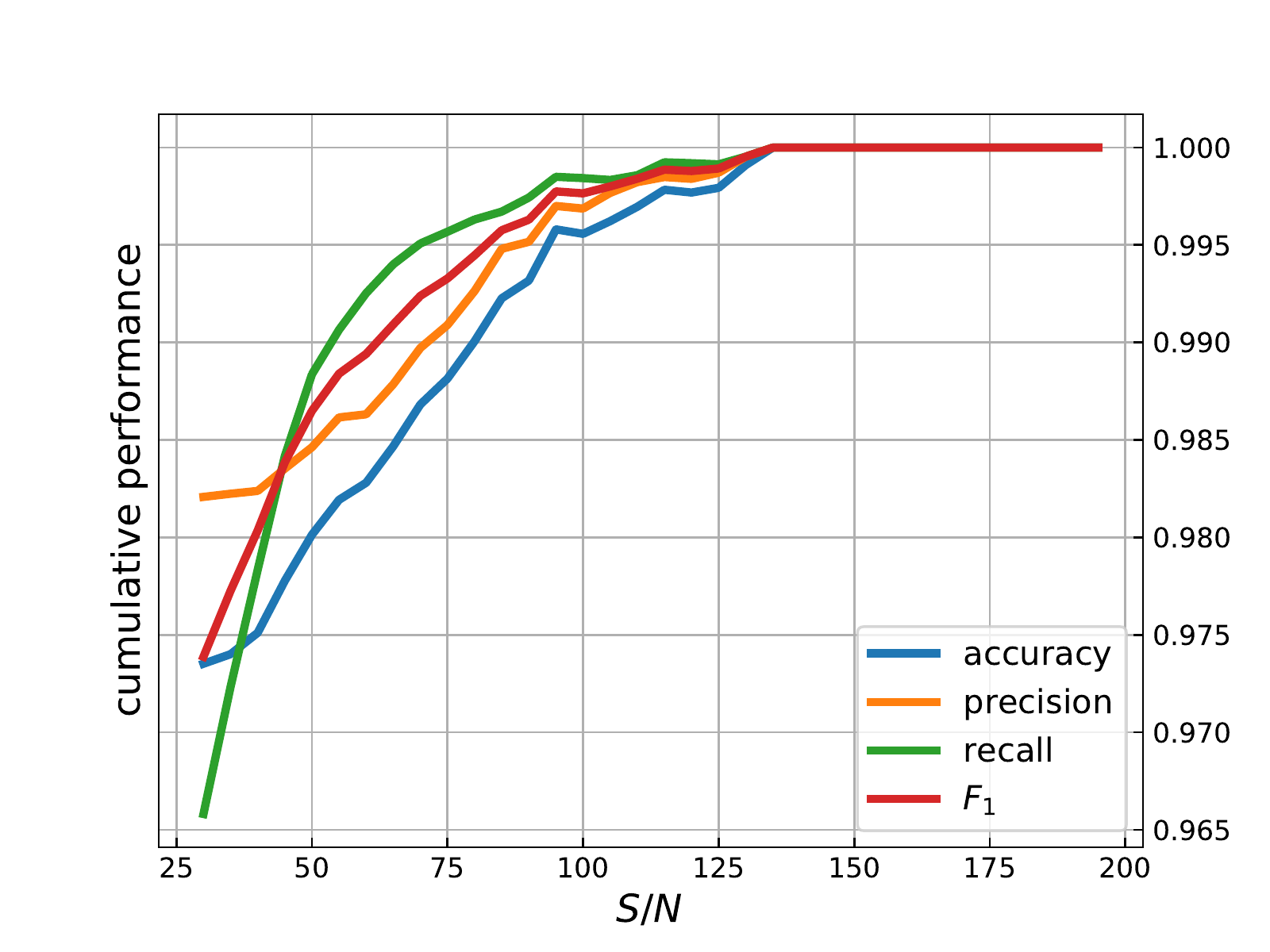}
\caption{Cumulative distributions of various performance metrics with respect to light-curve $S/N$ of our best model, measured on our explicit test set.
\label{fig:perf_seq}}
\end{figure}

Fig.~\ref{fig:perf_seq} shows the cumulative distribution of various performance metrics as a function of the light-curve $S/N$. As we can see, the performance is virtually $100\%$ for $S/N\gtrsim130$. We note that the dominance of RRab stars among the large-amplitude (and thus large $S/N$) variables in the studied period range greatly contributes to this figure. Beyond this limit, all metrics fall with decreasing $S/N$, as increasing noise washes out more and more characteristic features from the light curves. While the precision (purity) of the sample remains high including even the noisiest data dominated by non-RRab stars, recall (completeness) drops more drastically. Since the ratio of RRab and non-RRab stars in the training set varies with $S/N$, the $F_1$ measure, i.e., the harmonic mean of precision and recall, is the most informative metric in this test. The performance metrics measured on the test subset of $S/N>60$ are also shown in the last column of Table~\ref{tab:perf}. These values are very close to those measured by cross-validation, indicating that our model has excellent bias-variance tradeoff, thus it generalizes well on previously unseen data.

\begin{figure*}
\plottwo{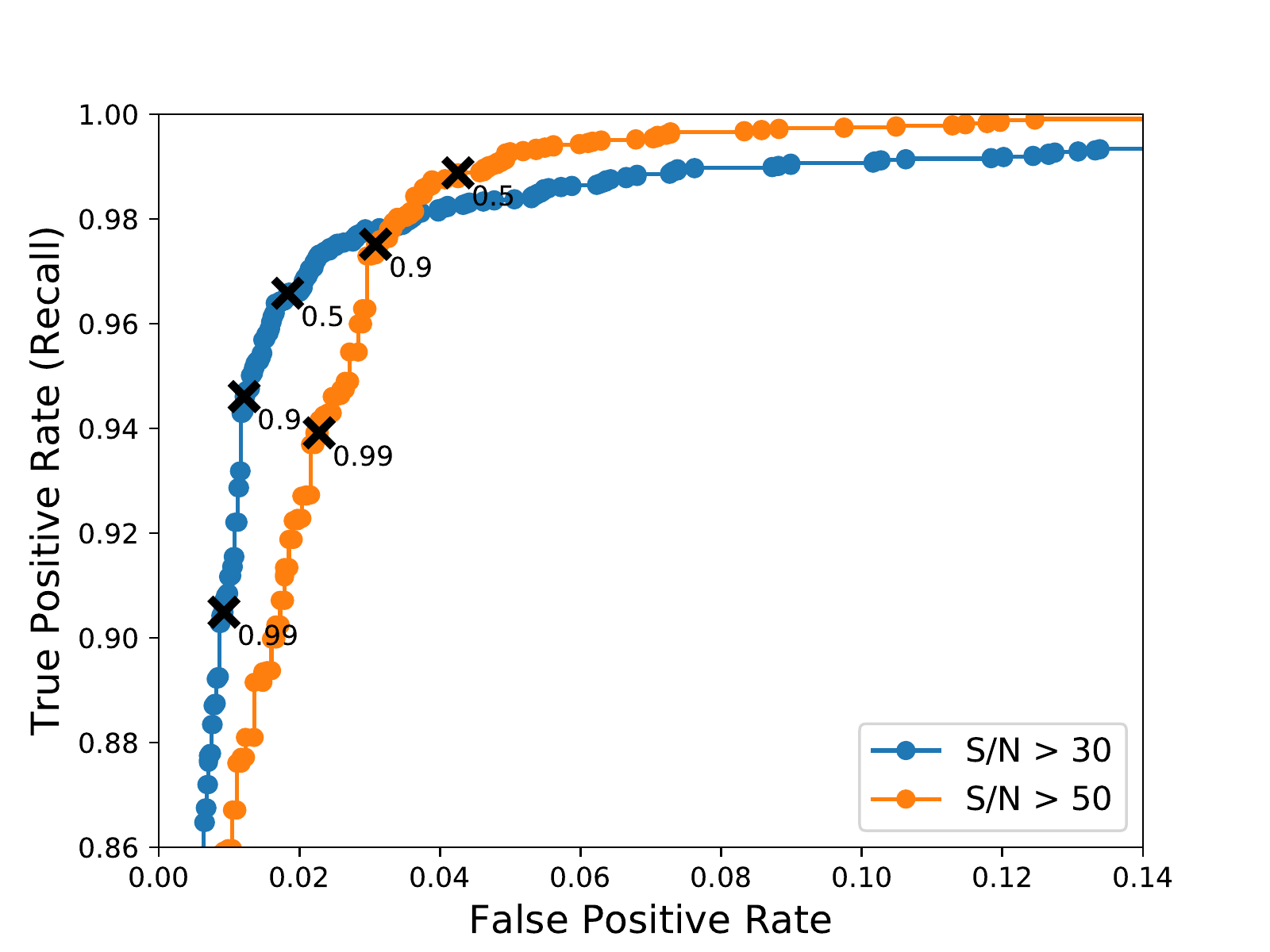}{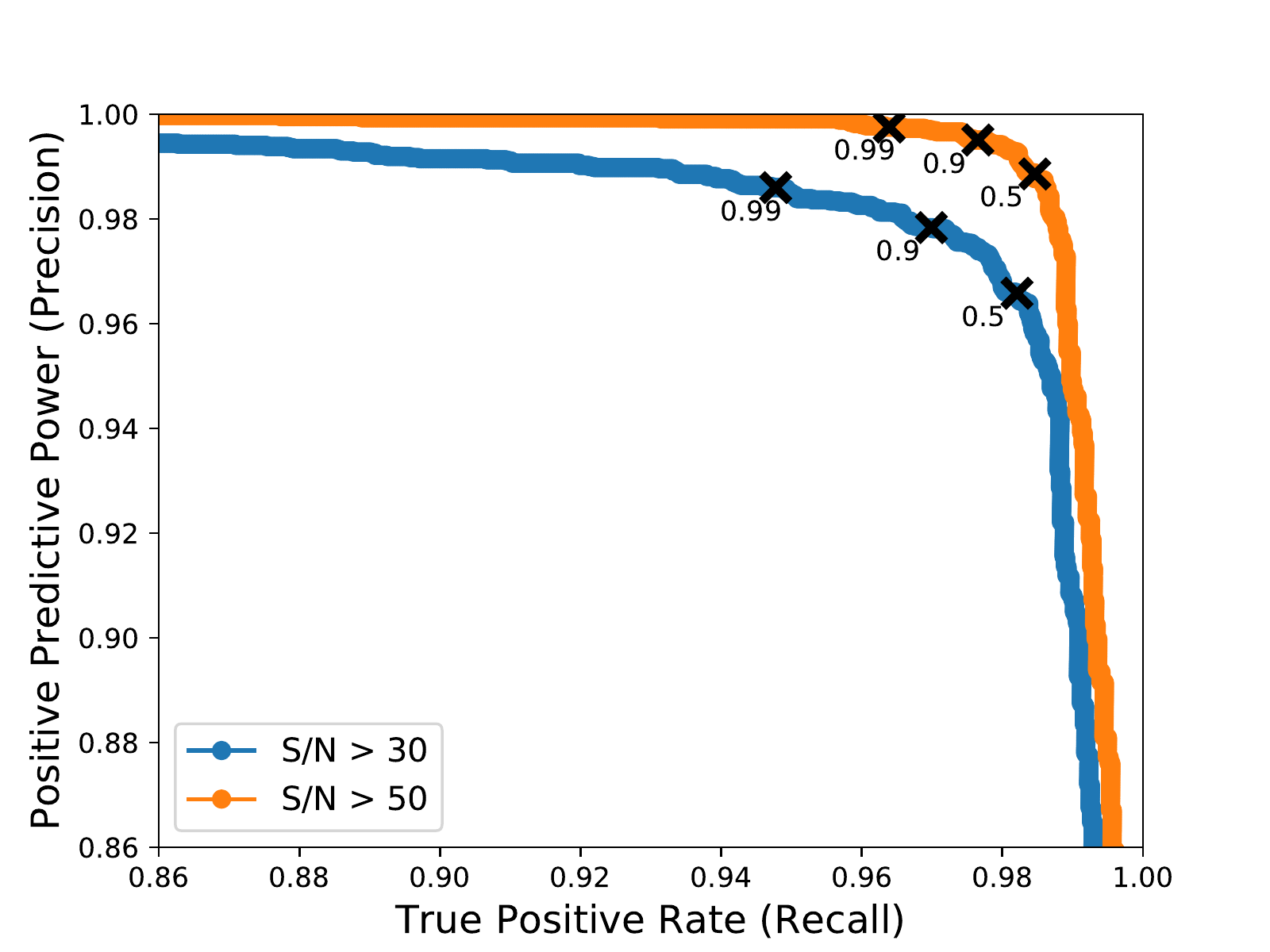}
\caption{
ROC (left) and precision-recall (right) curves of our best classifier measured on light curves in two different $S/N$ ranges in an explicit test set. Black crosses mark values for various decision boundaries annotated in the figure. Note the restricted ranges of the axes.
\label{fig:roc}}
\end{figure*}

Figure \ref{fig:roc} shows two additional performance diagnostics, namely the Receiver Operating Characteristic (ROC) curve and the precision--recall curve for our final classifier, computed for the test set. Both curves measure pairs of performance metrics as the decision boundary (i.e., the probability threshold) between the two classes is smoothly varied between 0 and 1. The area under the ROC curve is $0.993$ and $0.996$ computed for light curves with $S/N>30$ and $S/N>60$, respectively. The precision--recall curve can be characterized by the average precision score, which takes values of $0.993$ and $0.999$ for the same two ranges of light-curve $S/N$, respectively. The values of both metrics are very close to $1$, which would correspond to a perfect classifier. 

Based on these various tests, we can conclude that our RNN model supersedes the performance of the random forest RRab classifier of \citet{2016A&A...595A..82E} employed in our earlier near-IR census of RR~Lyrae stars along the southern disk, and is on par with  light-curve classifiers based on accurate and well-sampled optical photometric time series \citep[e.g.,][]{2019MNRAS.486.1907J}.

\section{The catalog of inner bulge RRab stars}\label{sec:catalog}

The RNN classifier developed in Sect.~\ref{sec:classification} was deployed on the set of periodic variable star candidates toward the inner bulge obtained according to Sect.~\ref{sec:observations}. First, each light curve was subjected to the regression procedure described in Sect.~\ref{lc_representation} in order to prepare the input sequences for classification.
Subsequently, the classifier was applied on those light curves that passed the following criteria:

\begin{eqnarray}
&&S/N \geq 30~; \label{eq:snr_sel}\\
&&N_{\rm ep} \geq 30~;\\
&&C_P \geq 0.8~;\\
&&C_{2P} \geq 0.8~;\\
&&\langle K_s \rangle \geq 12~{\rm mag};\\
&&A_{K_s}<1~{\rm mag}~;\label{eq:ampl_sel} \\ 
&&0.28\,{\rm d} \leq P \leq 0.98\,{\rm d}~; \label{eq:period_sel2}
\end{eqnarray}\label{target_set_criteria}

\noindent In total, $\simeq10^5$ light curves were subjected to the classification algorithm. Our RNN model gave a prediction of $\hat y >0.5$ for approximately $7\%$ of the target dataset. 

\begin{figure}
\includegraphics[width=0.5\textwidth]{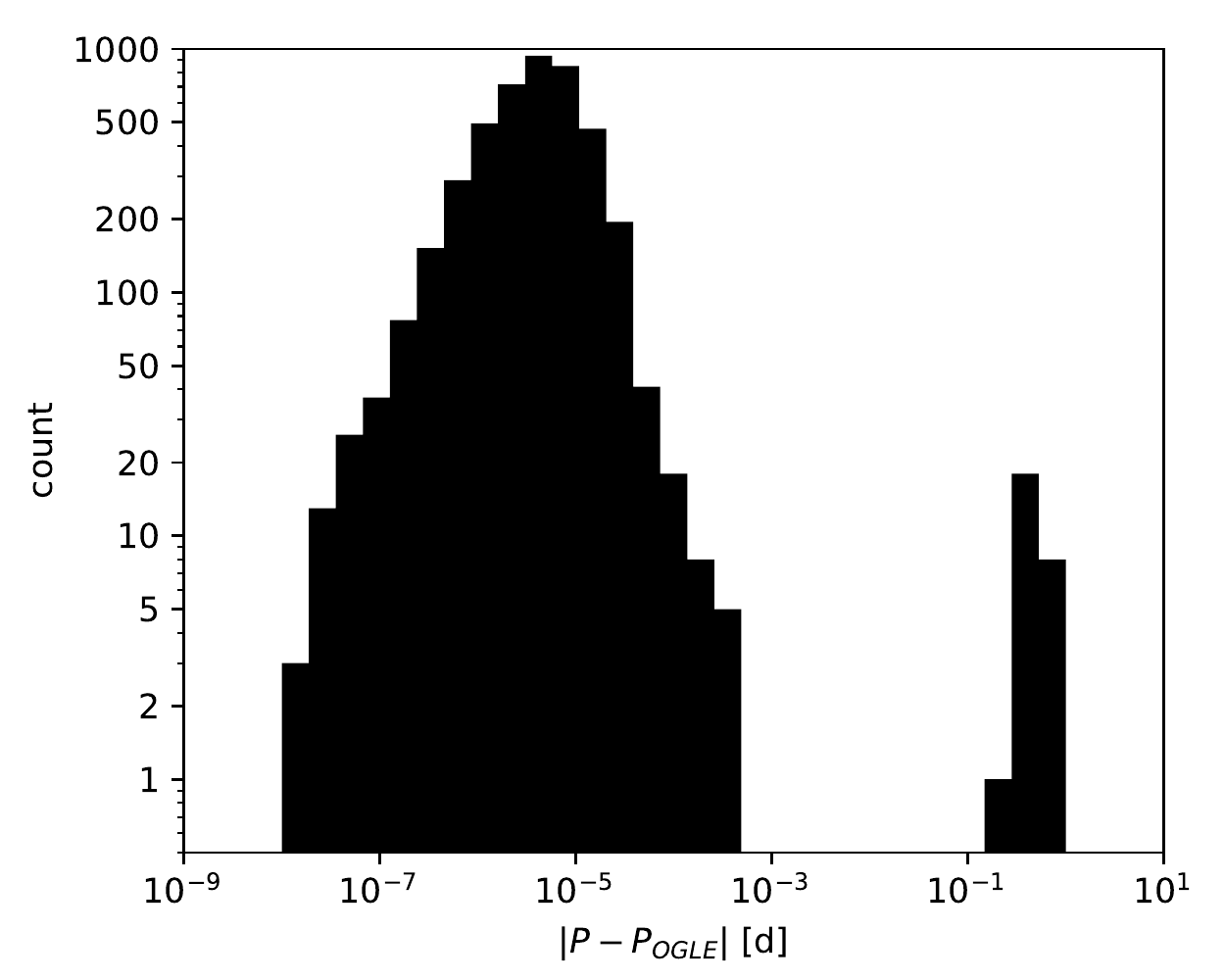}
\caption{Histogram of the absolute differences between the periods determined in this study ($P$) and determined by \citet[][$P_{\rm OGLE}$]{2019AcA....69..321S} for the OGLE RRab stars in our target dataset. Both axes have logarithmic scales.
\label{fig:period_diff_hist}}
\end{figure}

The presence of a large number of OGLE RRab stars in our target area allows us to test the period detection accuracy of our time-series analysis (Sect.~\ref{lc_representation}). Using a wide tolerance of $2''$ for angular separation, we cross-matched the positions of the OGLE-IV  RRab stars \citep{2019AcA....69..321S} with the catalog of periodic variable stars found within our target area and within the selection criteria in Eqs.~\ref{eq:snr_sel}--\ref{eq:ampl_sel}, but without any constraint on their periods derived from VVV data. Figure~\ref{fig:period_diff_hist} shows the histogram of the absolute differences between their periods derived from OGLE and VVV data. For all but a tiny fraction of objects, the period difference is on the order of $10^{-5}$---$10^{-6}$, thus we can conclude that the detection and classification efficiency of RRab stars in the VVV data is not significantly limited by the accuracy of the period determination.

The classification precision of the OGLE RRab stars in our target area is $97.2\%$. The corresponding 8734 objects in our target set that our RNN classified as RRab were previously discovered by OGLE. We note that $\sim80\%$ of the OGLE RRab stars misclassified by our RNN had $K_s$ light curves with $S/N<60$. In addition, among the OGLE catalog, we misclassified only 2 RRc stars and no RRd stars  as RRab. Both RRc stars have unusually long periods.

A similar cross-match with the Gaia DR2 catalog of RR~Lyrae stars \citep{2019A&A...622A..60C} resulted in 918 matches, 851 of which were also present in OGLE. Contrary to our classification, 38 of the matching Gaia objects were classified as either RRc or RRd by \citet{2019A&A...622A..60C}, 34 of which also contradict their OGLE classification as RRab.

\begin{figure*}
  \gridline{\fig{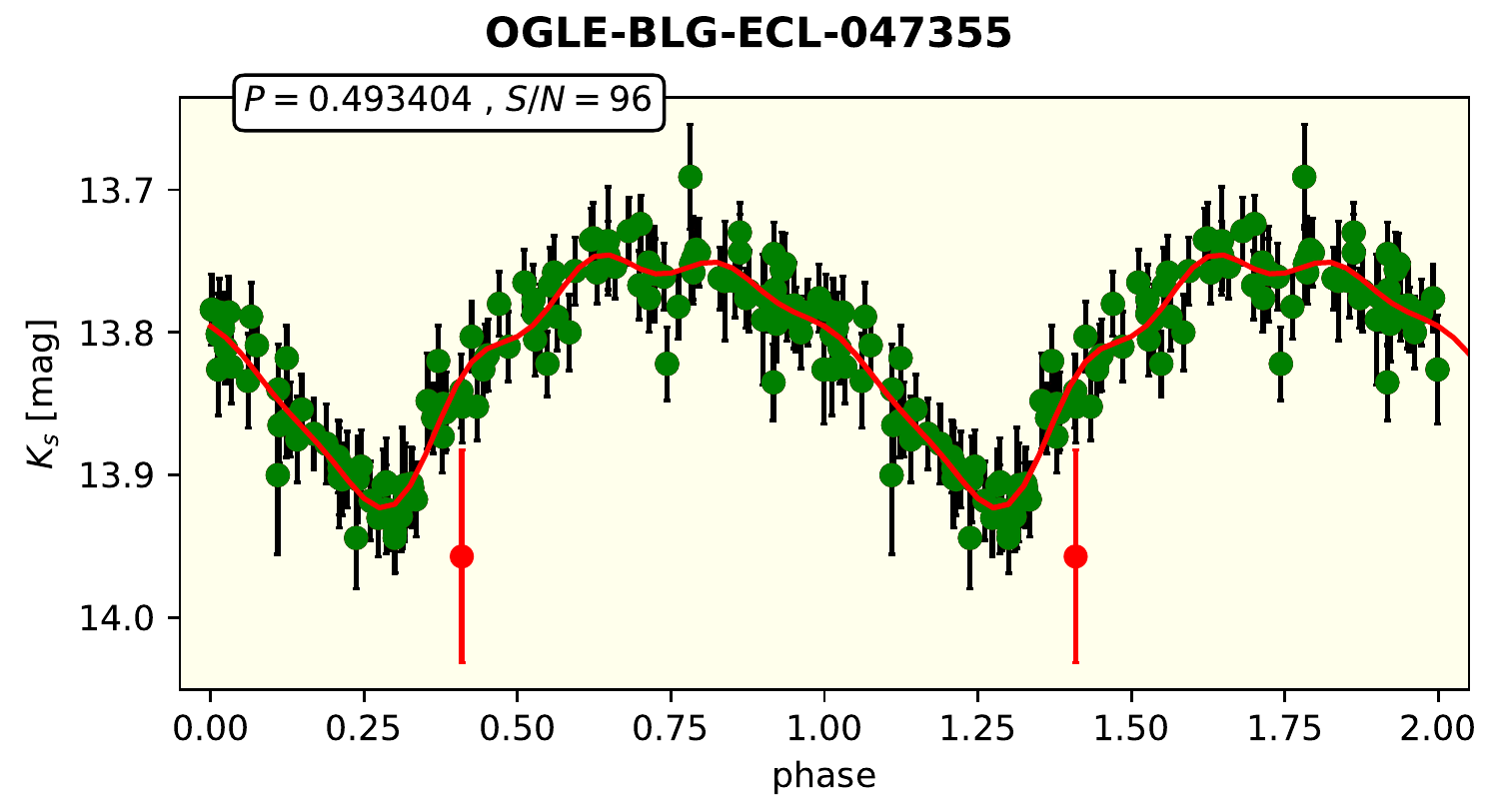}{0.48\textwidth}{}\fig{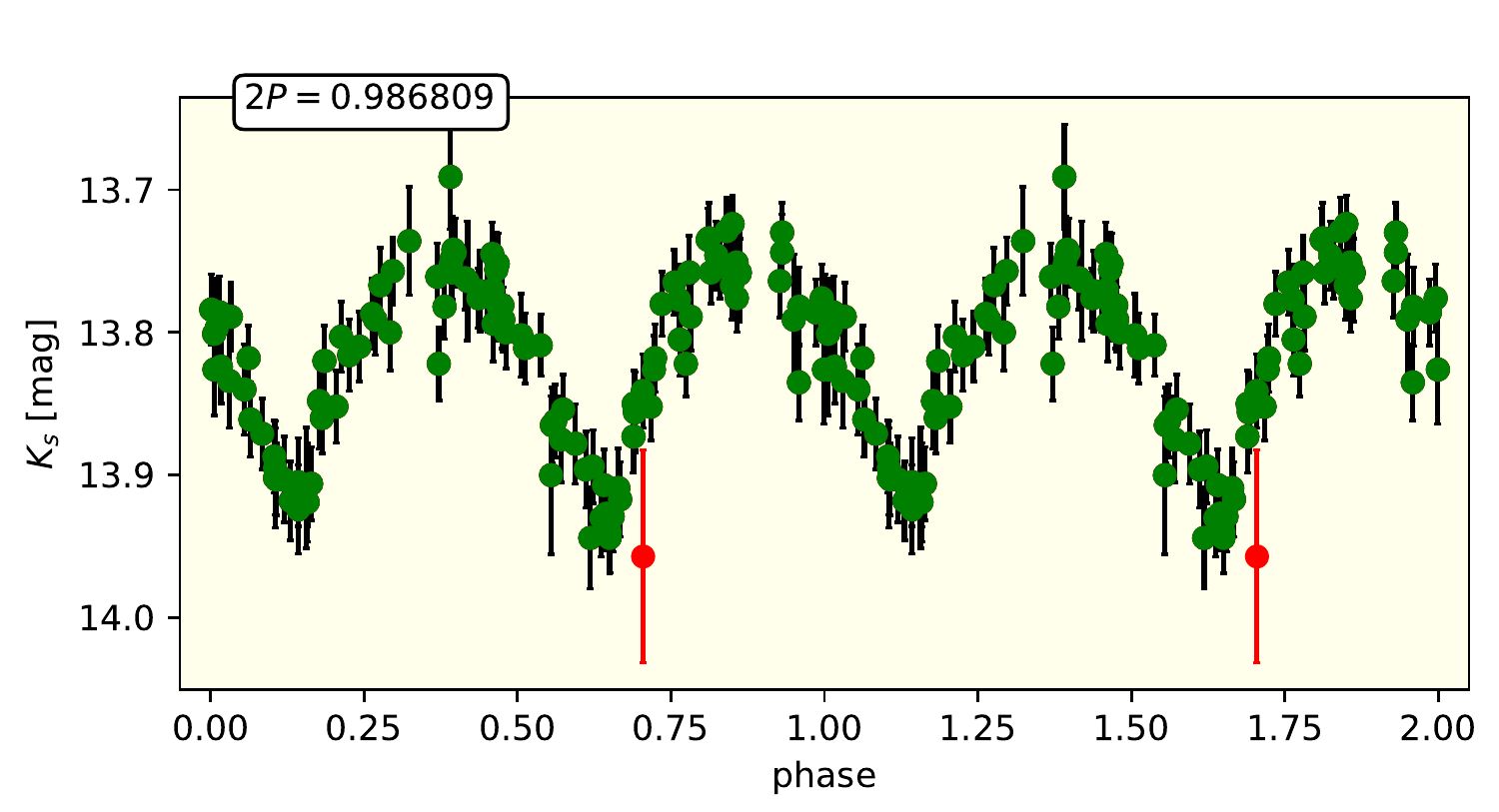}{0.48\textwidth}{}}
  \gridline{\fig{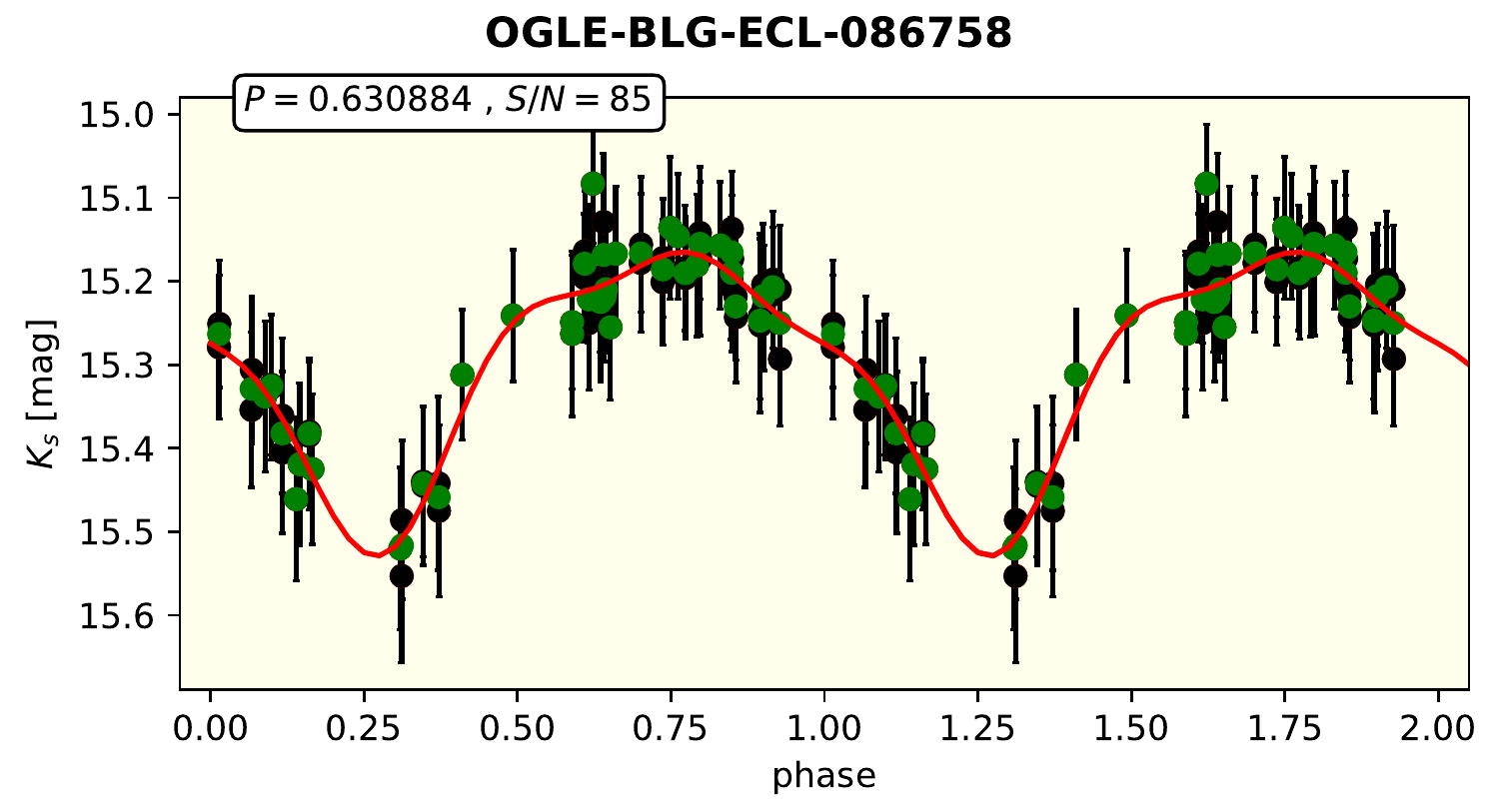}{0.48\textwidth}{}\fig{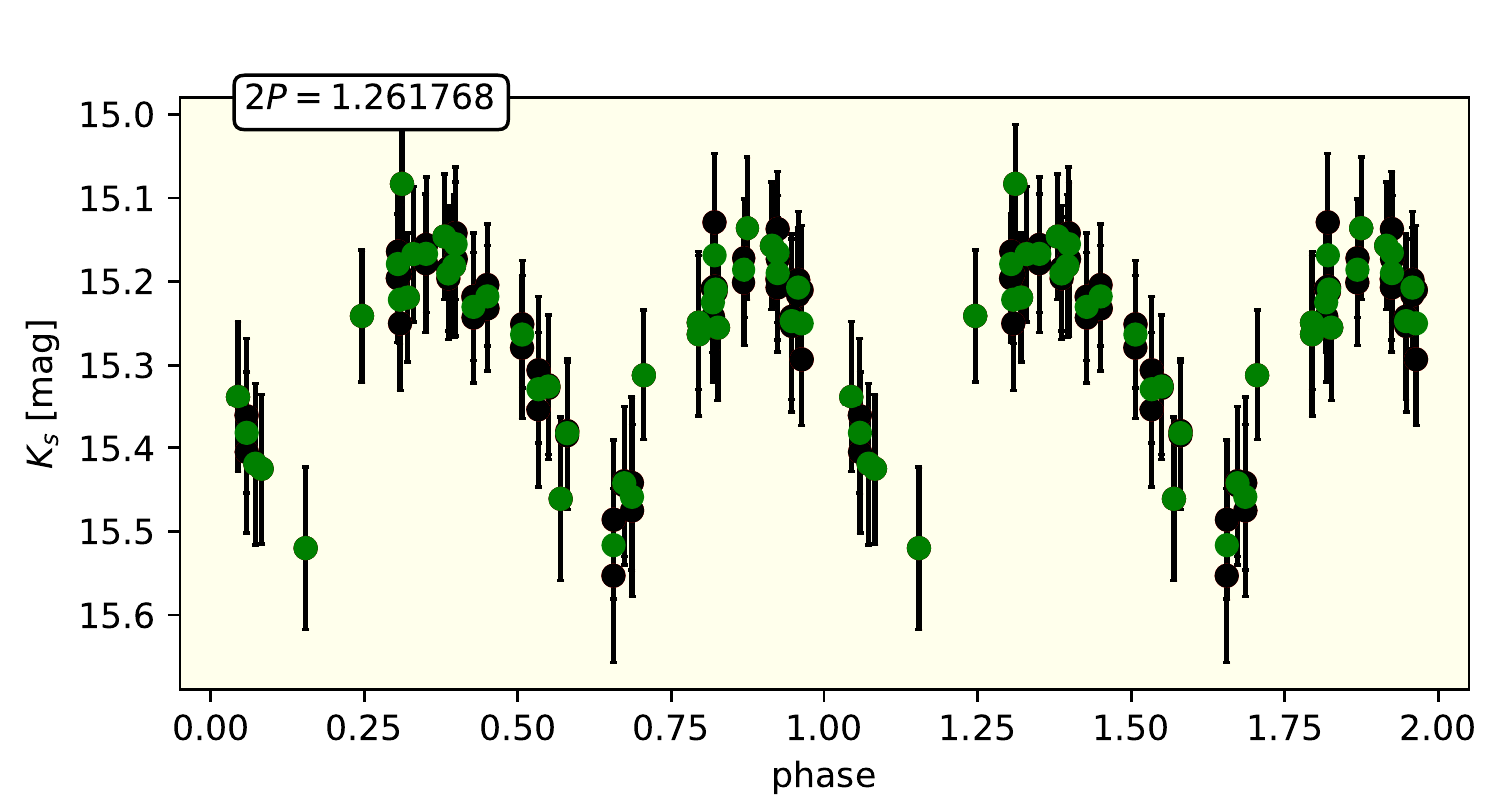}{0.48\textwidth}{}}
\caption{$K_s$-band VVV light curves of two eclipsing binaries known from the OGLE survey \citep{2016AcA....66..405S}, phase-folded with the period determined from VVV data (left), and with twice its value (right). The objects' identifiers are shown above the left panels. The notation is the same as in Fig.~\ref{fig:fit_example}.\label{fig:2examples}
}
\end{figure*}

The positions of the remaining objects in our sample were cross-matched with various catalogs of variable stars that could be confused with RRab stars due to their similar periods and amplitudes, since these were not {\em a priori} removed from our target set. These include the OGLE catalogs of classical, type II, and anomalous Cepheids \citep{2017AcA....67..297S}, eclipsing and ellipsoidal binary systems \citep{2016AcA....66..405S}, as well as other catalogs of Galactic classical Cepheids \citep[see][and references therein]{2019Sci...365..478S}. We found no match with any type of known Cepheid. However, we found 173 matches with the  catalog of binary stars, corresponding to a $\sim 2\%$ false discovery rate (defined as $N_{\rm fp}/(N_{\rm tp} + N_{\rm fp})$) if we account for the difference between the spatial coverage between the \citet{2016AcA....66..405S} catalog and our target area. For most of these stars, our analysis returned half of their true periods, in accordance with our findings in Sect.~\ref{lc_representation}. Figure~\ref{fig:2examples} shows two examples for binary stars misclassified as RRab by our RNN.

\begin{figure}
  \gridline{\fig{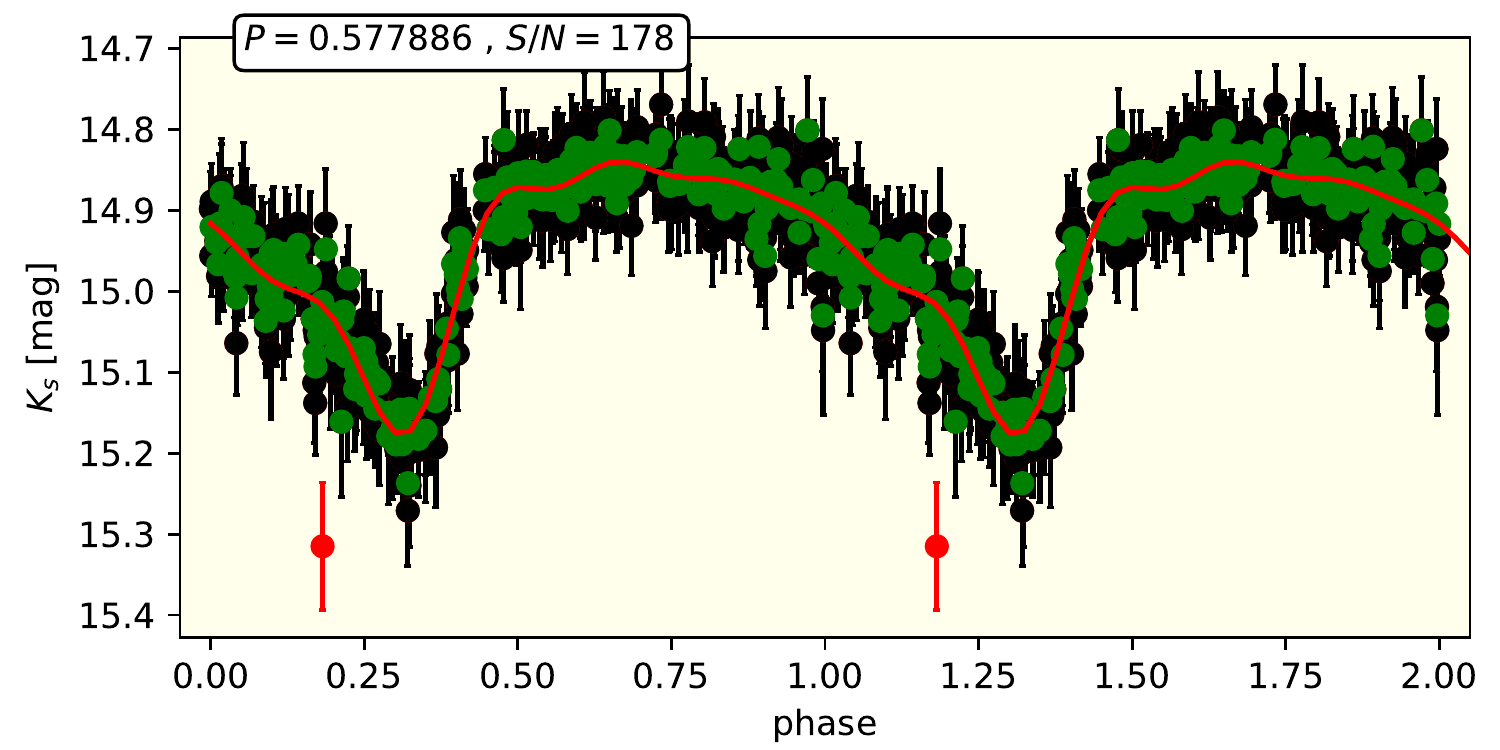}{0.48\textwidth}{}}
  \gridline{\fig{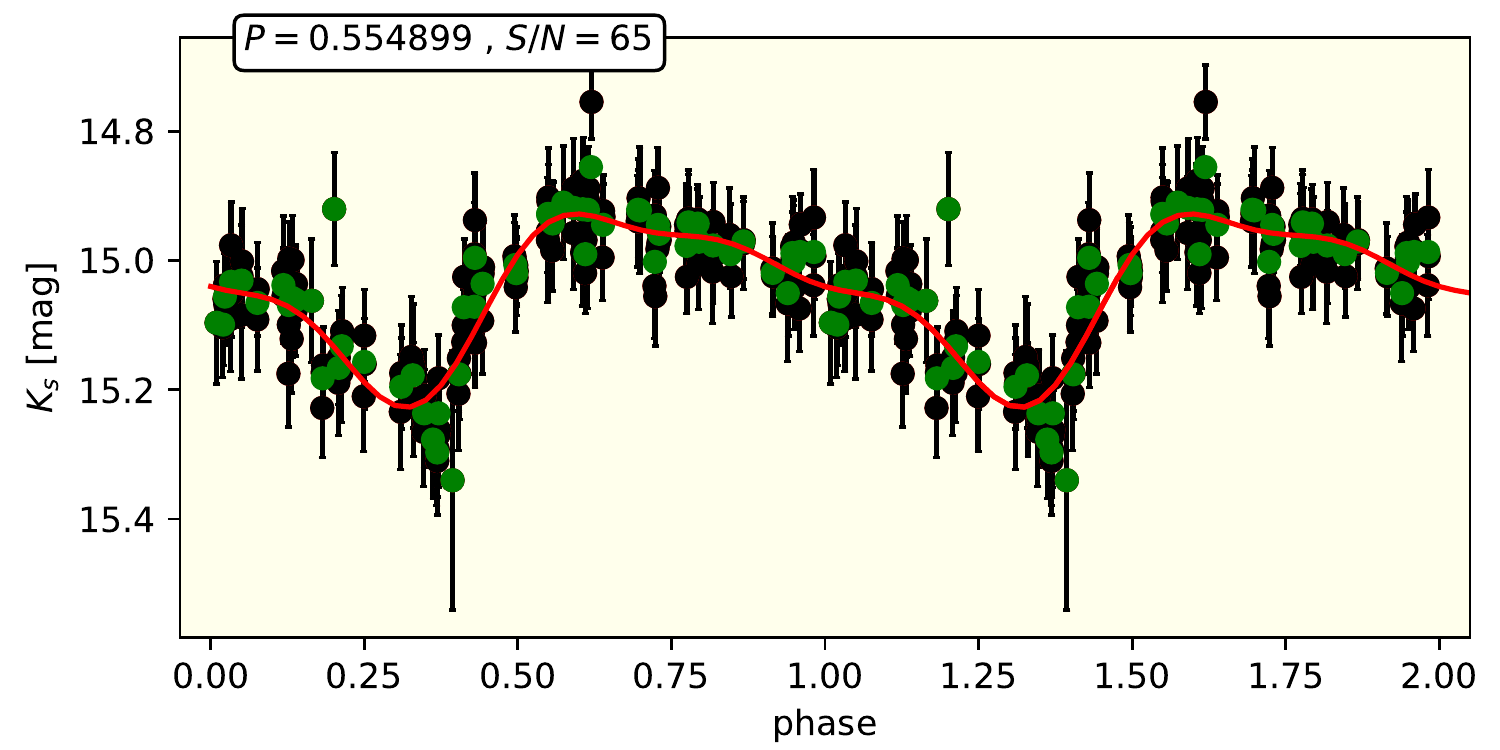}{0.48\textwidth}{}}
  \gridline{\fig{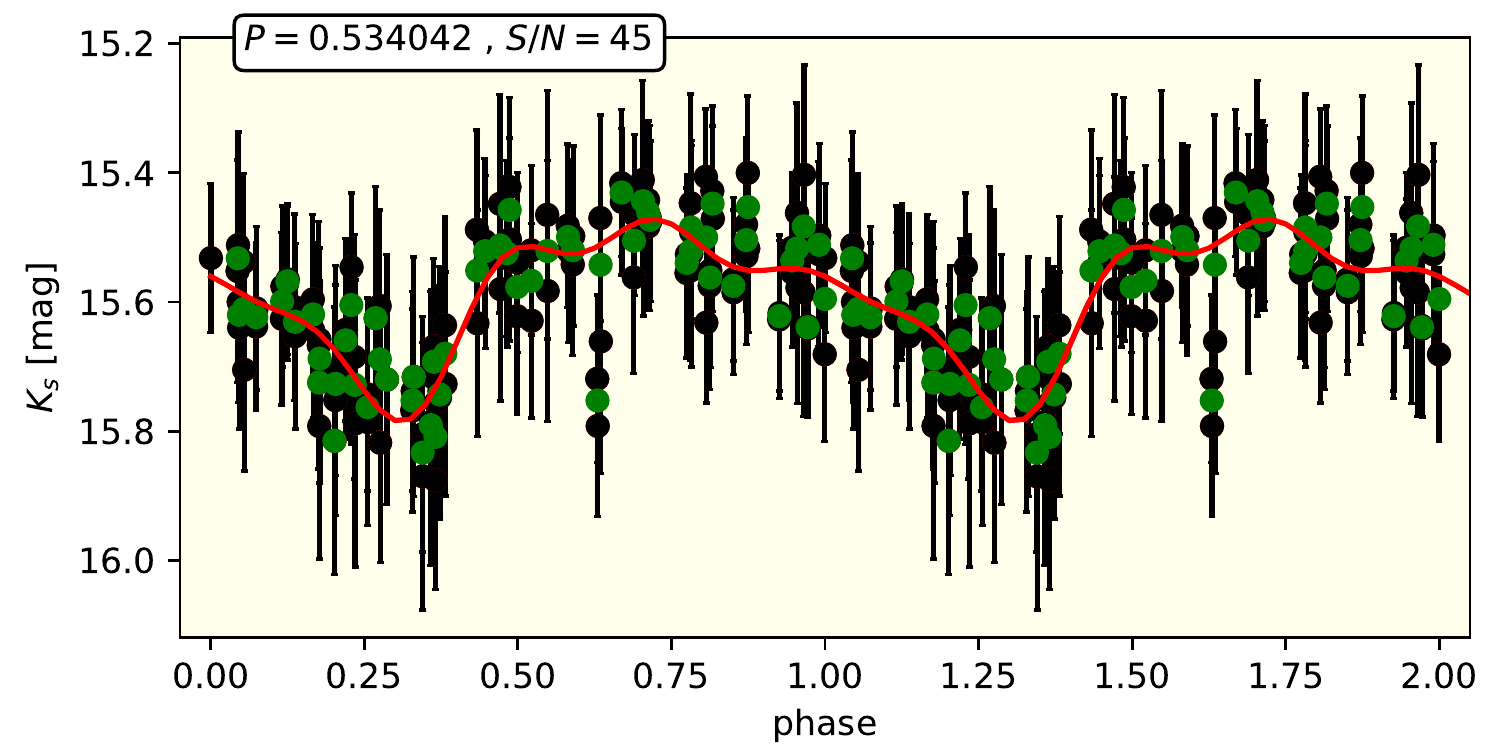}{0.48\textwidth}{}}
\caption{From top to bottom: phase-folded VVV $K_s$-band light curves of objects 12788, 10784, and 3929 from our final sample of RRab stars using the same notation as in Fig.~\ref{fig:fit_example}. Their periods and $S/N$ values are shown in the figure headers.
\label{fig:3examples}
}
\end{figure}

After the removal of the misclassified stars, our {\em final sample} contains 4447 objects classified as RRab stars, including the 102 visually classified objects found in the training and test sets (see Sects.~\ref{subsec:training_set} and \ref{subsec:performance}). Figure~\ref{fig:3examples} shows the light curves of three hitherto unknown objects in our final list, pertaining to different levels of $S/N$.

\begin{figure}
\includegraphics[width=0.48\textwidth]{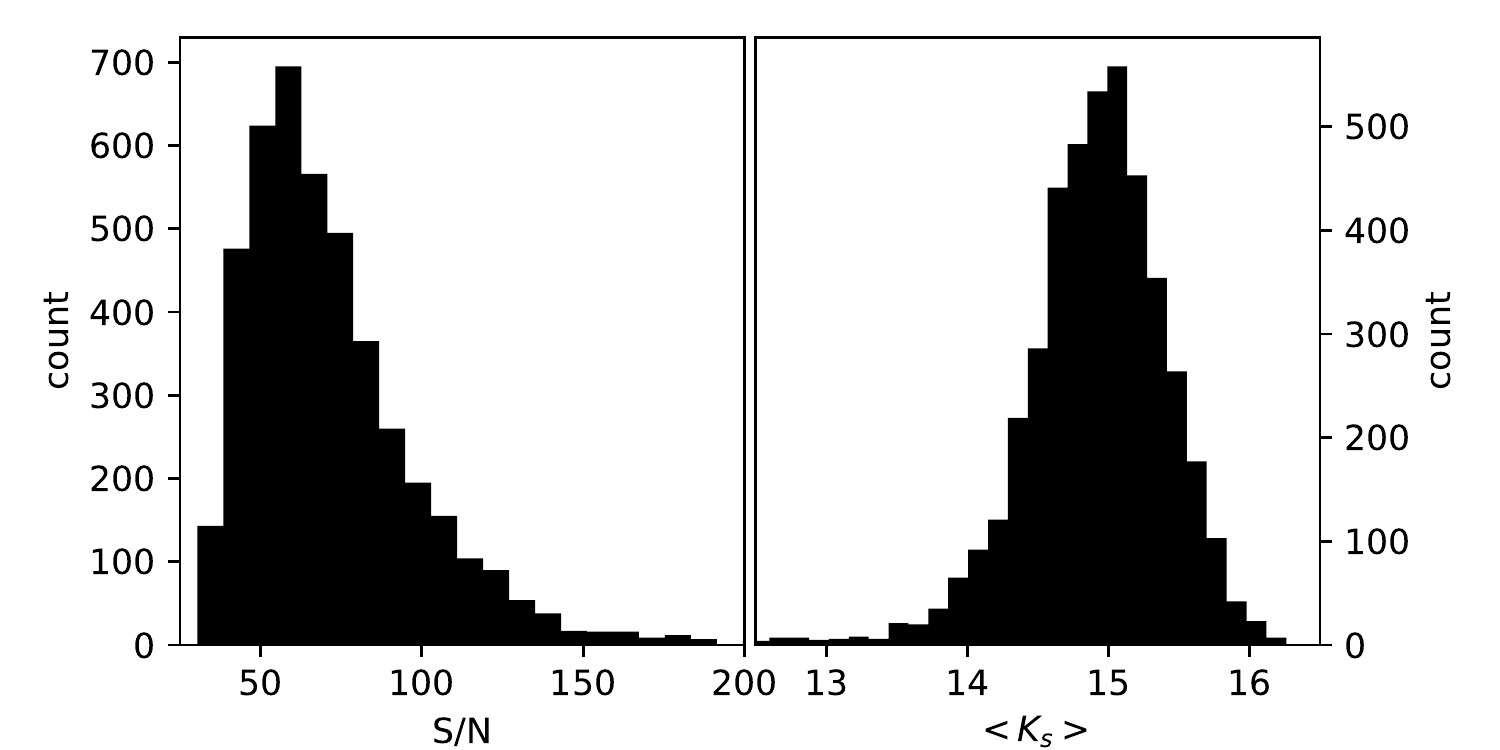}
\caption{Left: histogram of the $S/N$ values of the $K_s$ light curves of the objects in our final sample. Right: histogram of the mean apparent $K_s$ magnitudes of the same objects.
\label{fig:2_hist}}
\end{figure}

\begin{figure}
\includegraphics[width=0.5\textwidth]{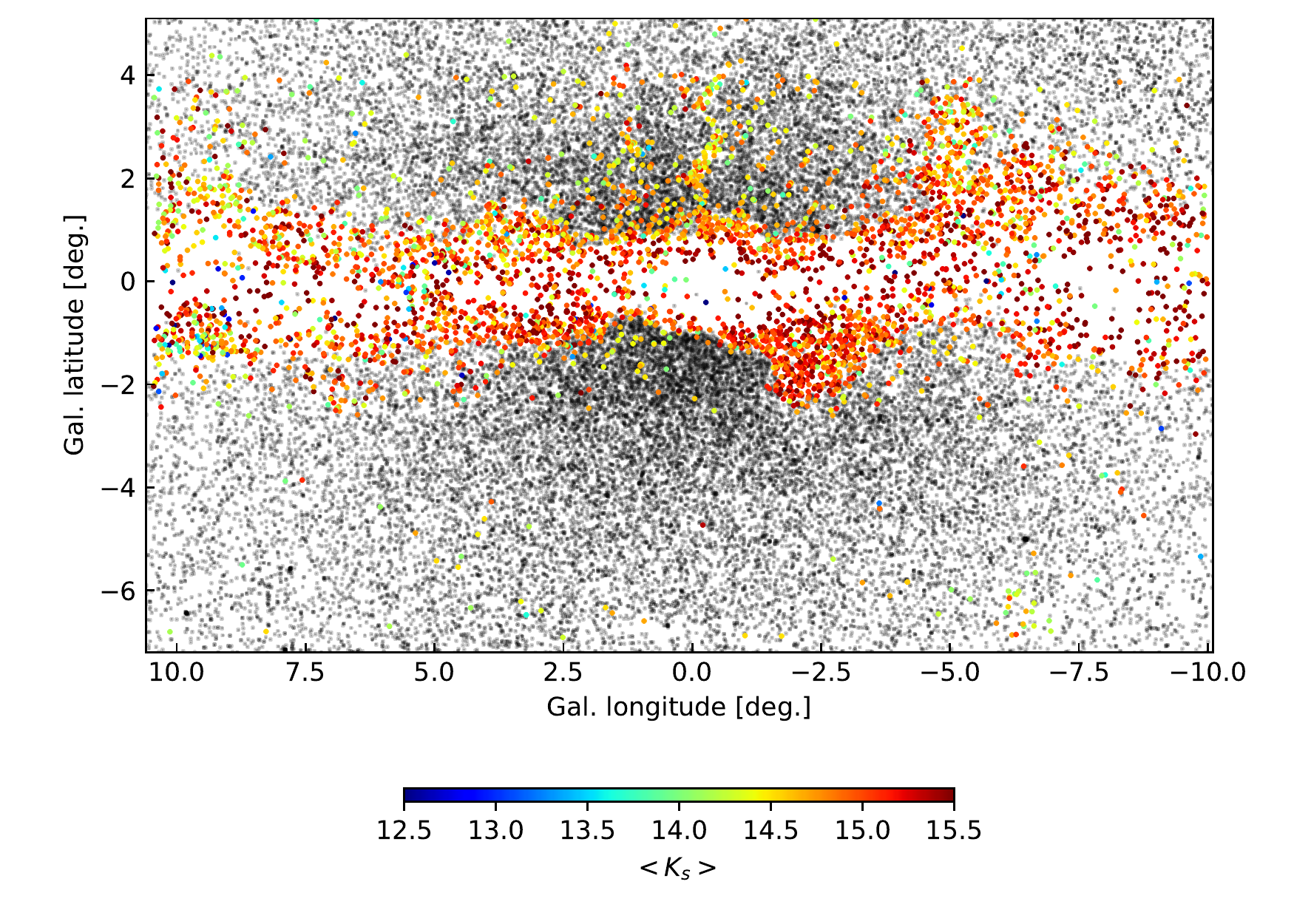}
\caption{Celestial distribution of the final sample of RRab stars detected in this study, shown in the Galactic coordinate system. Their mean apparent $K_s$ brightnesses are color-coded. Gray dots show the distribution of RRab stars known from the OGLE-IV survey and Gaia DR2. 
\label{fig:lb_map_newrr_kmean}}
\end{figure}

Figure~\ref{fig:2_hist} shows the histograms of the $S/N$ and the mean $K_s$ magnitudes of our final RRab sample, while Fig.~\ref{fig:lb_map_newrr_kmean} displays their celestial distribution with their apparent brightnesses color-coded. Our sample extends the census of RRab stars to the close proximity of the Galactic mid-plane where bulge RR~Lyrae stars are beyond the limiting magnitude of optical surveys due to interstellar extinction, and it also fills the gaps in the celestial coverage in the OGLE-IV survey. We note that new RRab stars following diagonal streaks are located in elongated gaps parallel with the ecliptic in the OGLE survey's coverage.

\begin{figure}
\includegraphics[width=0.48\textwidth]{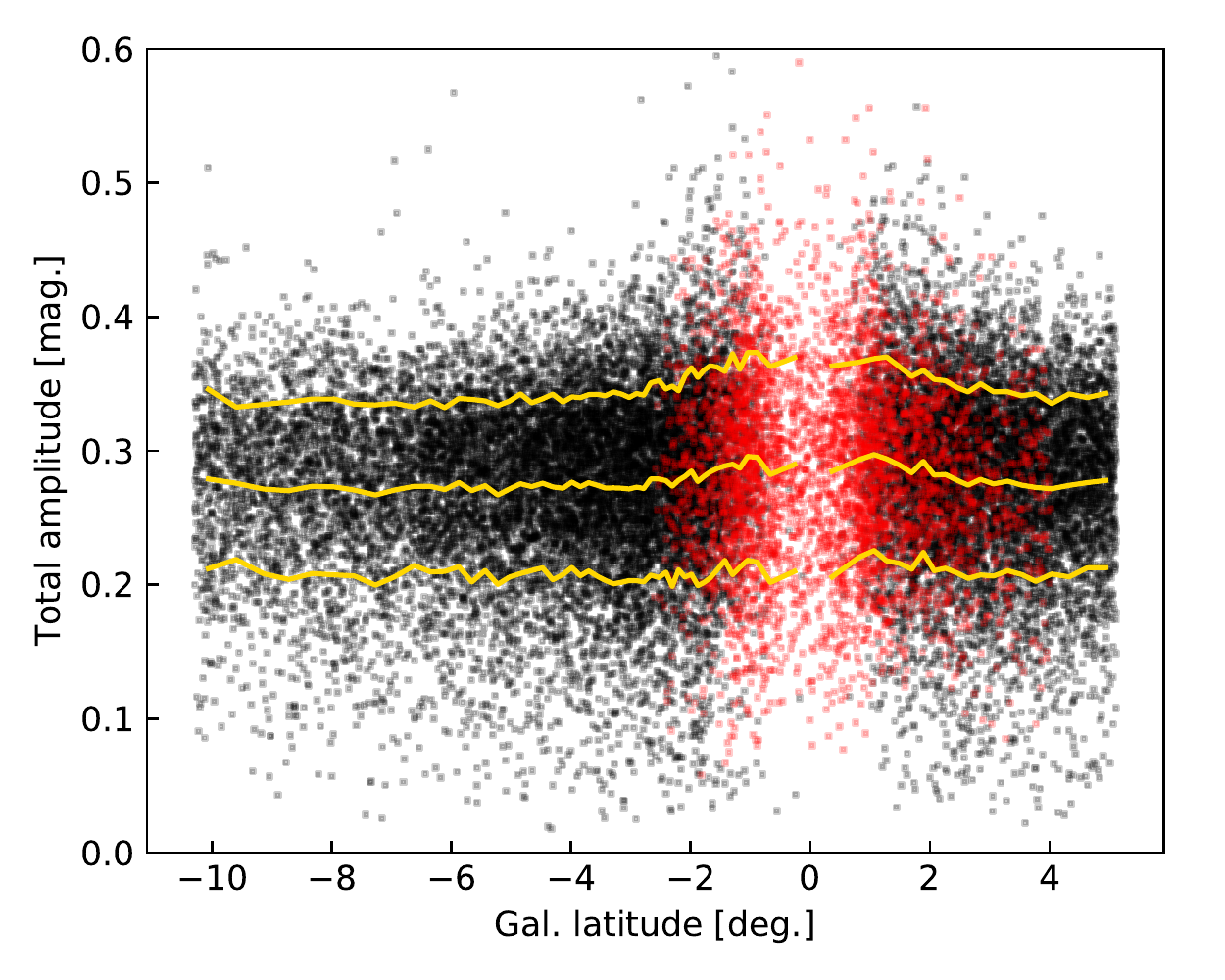}
\caption{Total (peak-to-valley) amplitudes of the $K_s$-band light curves of bulge RRab stars as a function of the Galactic latitude. Black: OGLE-IV RRab stars detected by the VVV survey; red: RRab stars in our final sample. The yellow lines denote the mean (middle line) and the mean $\pm$ the standard deviation (top and bottom lines) computed in latitude bins comprising a constant number of data points.
\label{fig:b_amp}}
\end{figure}

The drop in the number density of our sample toward some inner-bulge sight-lines follows the distribution of interstellar extinction \citep[e.g.,][]{2012A&A...543A..13G}, which, in spite of using the $K_s$ band, can obviously still push some RR~Lyrae apparent magnitudes beyond the detection limit of our analysis. Another factor that can lead to a local decrease in the sample's completeness is extreme source crowding. In this context, it is necessary to investigate the effect of point-source blending on our data. Figure~\ref{fig:b_amp} shows the $A_{K_s}$ total amplitudes resulting from our regression procedure (see Sect.~\ref{lc_representation}) for the objects in our final sample, in comparison with those of the OGLE-IV RRab stars with counterparts in the VVV survey, as a function of Galactic latitude. Since the point-source density, and thus the crowding steeply increases with latitude, in case of significant blending, the amplitudes would show a decreasing trend with decreasing angular distance from the mid-plane, as a result of the corresponding decrease in the relative flux variation of the objects due to the flux contamination by blending objects. Such effect cannot be observed; on the contrary, the total amplitudes show a slight increase toward the Galactic plane, probably caused by the increasing noise in the light curves. We conclude that our aperture optimization procedure significantly mitigates the effect of source crowding, and that the photometry of our RRab sample is not significantly impaired by blending.

\begin{figure*}
\plottwo{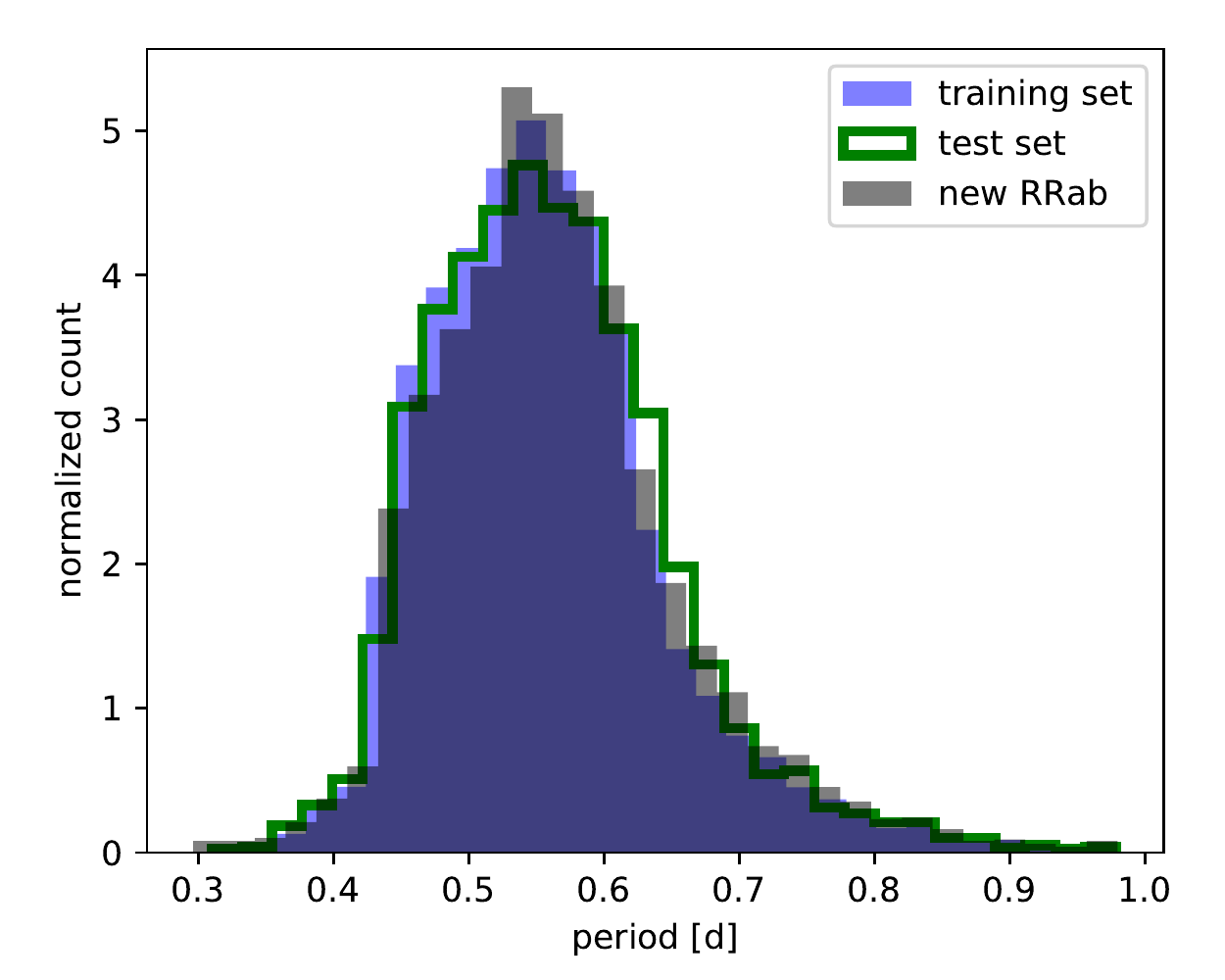}{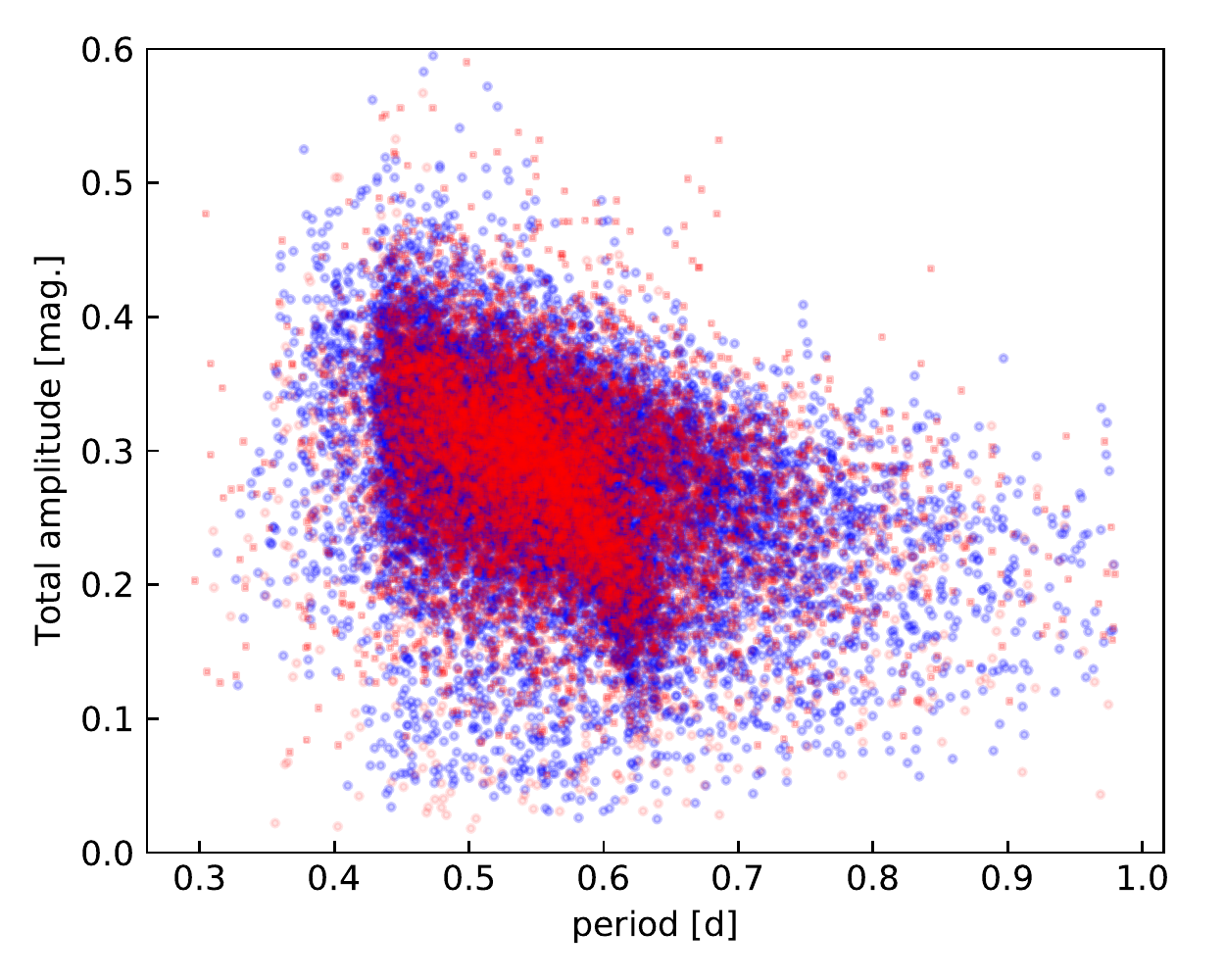}
\caption{Left: Histogram of the periods of RRab star in our training, test, and newly detected RRab stars. Right: $K_s$-band amplitude vs period diagram of the RRab stars in our training+test sets (blue) and of the newly identified RRab stars (red).
\label{fig:per_amp}}
\end{figure*}

The period and period-amplitude distributions of the RRab stars in our final sample are shown in Fig.~\ref{fig:per_amp}, in comparison with the objects in our training and test sets. It is immediately evident that all samples share the same distribution, indicating the high purity of our inner bulge RRab catalog. We emphasize that while short- ($P\lesssim0.45$\,d) and long-periodic ($P\gtrsim0.7$\,d) objects are naturally underrepresented in our training set, the recall of our classifier does not vary significantly with period, which is a remarkable property of the RNN architecture.

The target area of our RRab census overlaps with the $100'$-radius cone  around the Galactic Center previously searched by \citet{2018ApJ...863...79C} for RR~Lyrae stars. We cross-matched the positions of our final sample with their catalog of 488 objects not already included in the OGLE-IV and Gaia DR2 RR~Lyrae catalogs, resulting in 133 common objects. The remaining 4314 RRab stars in our final sample are new discoveries to our best knowledge. We note that the analysis of \citet{2018ApJ...863...79C} was based on custom PSF photometry, allowing slightly higher $S/N$ and a deeper variability search toward the Galactic Center. Unfortunately, we cannot confirm the classification of the rest of the objects in their catalog, in the absence of published light curves.

Table~\ref{tab:catalog} lists the names, coordinates, periods, and basic $K_s$ photometric properties, classification probabilities and cross-identifications of all RRab stars in our final sample ordered by right ascension, also including the data of previously catalogued RR~Lyrae stars in our target area. Based on the $S/N$ and the probability given for each object, specific subsamples for various further applications of the objects can be easily drawn with an appropriately tuned balance between precision and recall. Finally, in Table~\ref{tab:tseries}, we provide the $J$-, $H$-, and $K_s$-band photometric time-series of all objects listed in Table~\ref{tab:catalog}.

\begin{deluxetable*}{cccccccccccc}
\tablecaption{Coordinates and basic photometric properties of the RRab stars \label{tab:catalog}}f
\tablehead{
\colhead{Name} & \colhead{RA (J2000.0)} & \colhead{DEC (J2000.0)} &
\colhead{period~[d]} & \colhead{$\langle K_s \rangle$} & \colhead{$A_{K_s}$} &
\colhead{aper.\tablenotemark{a}} & \colhead{$S/N$} & \colhead{prob.\tablenotemark{b}} & 
\colhead{Gaia SourceID} & \colhead{OGLE ID}\tablenotemark{c} & \colhead{flag}\tablenotemark{d}
}
\startdata
     1   &   17:03:50.24  & -34:52:36.8 &  0.604904  & 14.328 &  0.199 &   2  &    121.7 &  0.998 &  \dots                               &  42687   &       \dots      \\
     2   &   17:03:54.70  & -34:49:50.2 &  0.586816  & 15.164 &  0.283 &   2  &    118.1 &  1.000 &  5978072173660351744 &  42717   &       \dots      \\
     3   &   17:04:50.24  & -34:56:26.6 &  0.683041  & 13.922 &  0.255 &   3  &    170.7 &  1.000 &  5978022180236791552 &  42978   &       \dots      \\
     4   &   17:04:50.77  & -34:34:51.3 &  0.491694  & 15.075 &  0.270 &   2  &    102.2 &  1.000 &  5978083001389214464 &  42980   &       \dots      \\
     5   &   17:04:52.75  & -35:09:08.5 &  0.450633  & 14.801 &  0.296 &   3  &    108.9 &  1.000 &  5978005966747456256 &  42993   &       \dots      \\
\enddata
\tablenotetext{a}{Optimal aperture (see Sect.~\ref{lc_representation}).}
\tablenotetext{b}{Classification probability ($\hat y$).}
\tablenotetext{b}{The identifier following `OGLE-BLG-RRLYR-' in \citet{2019AcA....69..321S}.}
\tablenotetext{d}{`v': visual classification (see Sects.~\ref{subsec:training_set} and \ref{subsec:performance}); `C18': object listed by \citet{2018ApJ...863...79C}}
\tablecomments{This table is available in its entirety in machine-readable form.}
\end{deluxetable*}

\begin{deluxetable}{cDlccDD}[]
\tablecaption{Photometric time-series of the RRab stars. \label{tab:tseries}}
\tablehead{
\colhead{Name} & \multicolumn2c{HJD-2400000} & \colhead{filter} & pawprint\tablenotemark{a} & chip & \multicolumn2c{mag.} & \multicolumn2c{mag.err.}}
\decimals
\startdata
1  & 55305.800250 & $K_s$ & 111444 & 4 &  14.434 & 0.019 \\
1  & 55305.807440 & $K_s$ & 111469 & 4 &  14.418 & 0.021 \\
1  & 55407.656280 & $K_s$ & 149730 & 1 &  14.254 & 0.021 \\
1  & 55437.613440 & $K_s$ & 172155 & 4 &  14.405 & 0.022 \\
1  & 55444.519520 & $K_s$ & 175854 & 1 &  14.253 & 0.022 \\
\enddata
\tablenotetext{a}{Identifier of the corresponding VISTA pawprint (single detector image stack), identical to the value of the ``HIERARCH ESO DET EXP NO'' ESO header keyword in the CASU photometric catalog.}
\tablecomments{This table is available in its entirety in machine-readable form.}
\end{deluxetable}

\newpage
\section{Summary}\label{sec:summary}

We leveraged near-IR photometry from the VVV survey to search for RRab stars in a $\sim121.5$ square-degree area toward the inner Galactic bulge. In order to separate the RRab stars from other point sources showing periodic signals, we trained a deep LSTM RNN classifier on VVV data, which is the first application of this neural network architecture for the near-IR light-curve classification of pulsating variable stars. Our classifier attained an $F_1$ score of better than $\sim97\%$ on an explicit test data set containing variable sources in a wide range of $S/N$ with a distribution that matches well our target data set, reaching $F_1\simeq99\%$ for $S/N>60$. Our model thus demonstrates for the first time that a classification performance similar to those reached by the current best optical light-curve classifiers can be reached for near-IR $K_s$ data by deep learning employing RNNs. Based on various performance estimates and the fraction of misclassified objects in the target dataset, we can estimate that the contamination rate in our final sample of RRab stars is a few percent.

Our search resulted in the discovery of 4314 bona fide RRab stars concentrated to $\vert b \vert \lesssim 1^\circ$, where most of these objects are beyond the detection limit of optical surveys. While this is a significant extension of the RR~Lyrae census to the most observationally inaccessible region of the sky, due to extreme levels of extinction and source crowding, part of the inner bulge RRab stars toward the Galactic Center and along the sight-lines of dust clouds in the close proximity of the mid-plane still remain uncovered. These areas can be further searched by exploiting near-IR data by, e.g., image subtraction techniques, as well as by mid-IR surveys. For the former, our RNN classifier will come in useful. The catalog of new inner bulge RRab stars published in this study can greatly contribute to the better understanding of the properties of interstellar extinction along the Galactic plane, as well as the 3-dimensional structure and kinematics of the oldest stellar populations of the inner Galaxy.


\acknowledgments
We would like to thank the anonymous referee for his/her valuable comments which greatly helped to improve this study.
Our results were based on observations collected at the European Southern Observatory under ESO programme 179.B-2002.
The authors were supported by the Deutsche Forschungsgemeinschaft (DFG, German Research Foundation) -- Project-ID 138713538 -- SFB 881 (``The Milky Way System'', subproject A03).


\vspace{5mm}
\facilities{ESO:VISTA}


\software{STIL \citep{2006ASPC..351..666T},  
                numpy \citep{numpy},  
                scipy \citep{scipy},  
                TensorFlow \citep{2016arXiv160304467A},  
                Keras \citep{keras} 
          }




\end{document}